\documentclass[12pt]{article}

\usepackage[numbers, sort&compress]{natbib}
\usepackage{fullpage}
\usepackage{epsfig}
\usepackage{latexsym,amsmath,amssymb,amsthm}
\usepackage{subfig}
\newcommand{\ceil}[1]{\left\lceil {#1} \right\rceil}
\newcommand{\floor}[1]{\left\lfloor {#1} \right\rfloor}

\def\maps11{\stackrel {1-1}{\longmapsto}}

\begin{document}

\title{Coupled IEEE 802.11ac and TCP Goodput improvement
using Aggregation and Reverse Direction}

\author{%
Oran Sharon
\thanks{Corresponding author: oran@netanya.ac.il, Tel: 972-4-9831406,
Fax: 972-4-9930525} \\
Department of Computer Science \\
Netanya Academic College \\
1 University St. \\
Netanya, 42365 Israel
\and
Yaron Alpert\\
Intel\\
13 Zarchin St.\\
Ra'anana, 43662, Israel\\
Yaron.alpert@intel.com
}


\date{}

\maketitle

\begin{abstract} 
This paper suggests a new model for the transmission
of Transmission Control Protocol (TCP)
traffic over IEEE 802.11 using
the new features of IEEE 802.11ac . The paper
examines a first step in this direction and as
such we first consider a single TCP connection, which is
typical in a home environment. 
We show that when the IEEE 802.11ac MAC is
aware of QoS TCP traffic, using Reverse Direction improves
the TCP Goodput in tens of percentages compared to the
traditional contention
based channel access. In an error-free channel this improvement
is $20\%$ while in an error-prone channel the improvement
reaches $60\%$, also using blind retransmission of frames.
In our operation modes we also assume the use in Two-Level aggregation
scheme, the Automatic Repeat-Request (ARQ)
protocol of the IEEE 802.11ac MAC layer
and also assume the data rates
and the four Access Categories defined in this standard.
\end{abstract}

\bigskip

\noindent
\textbf{Keywords}: 802.11ac; TCP; Aggregation; Reverse Direction; Goodput;

\renewcommand{\baselinestretch}{1.3}
\small\normalsize


\section{Introduction}

\subsection{Background}

\indent
The latest IEEE 802.11-REVmc Standard (WiFi), created and maintained by 
the IEEE LAN/MAN Standards Committee (IEEE 802.11)~\cite{IEEEBase1},
that embedded and updated the IEEE 802.11ac amendment,
is currently the most
effective solution within the range of Wireless Local
Area Networks (LAN). Since its first release in 1997,
the standard provides the basis
for Wireless network products
using the WiFi brand, and has since been improved upon
in many ways. One of the main goals of these improvements
is to optimize the Throughput of the MAC layer, and to improve
its Quality-of-Service (QoS) capabilities.

To fulfill the promise of increasing IEEE 802.11 performance and
QoS capabilities, 
and to effectively support more client devices on a network,
the IEEE 802.11 working group introduced the fifth 
generation in IEEE 802.11 networking standards;
the IEEE 802.11ac amendment, 
also known as Very High Throughput (VHT)~\cite{IEEEac, IEEEBase1}. 
IEEE 802.11ac
is intended to support fast, high quality data streaming and nearly
instantaneous data syncing and backup to notebooks, tablets
and mobile phones. The IEEE 802.11ac final version, 11ac-2013, released
in 2013~\cite{IEEEac},
leverages new
technologies to provide improvements over previous generation, i.e.
IEEE 802.11-2012~\cite{IEEEBase} .
Both versions are now included in IEEE 802.11-REVmc~\cite{IEEEBase1}
which will be published as IEEE 802.11 - 2016.

The IEEE 802.11ac amendment~\cite{IEEEac}
improves the achieved Throughput
coverage and QoS capabilities, compared to
previous generations, by introducing improvements
and new features in the PHY and MAC layers. In the PHY layer,  
IEEE 802.11ac (VHT) 
continues the long-existing trend towards higher 
Modulation and Coding rates ( 256 QAM 5/6 modulation), working
in wider bandwidth channels ( up to 160 MHz ) and using 8 spatial
streams that enable higher spectral efficiency.

In the MAC layer IEEE 802.11ac includes many of the improvements
first introduced with IEEE 802.11e and IEEE 802.11n~\cite{IEEEBase},
also known as 
High Throughput (HT). 
Integrated with the following two key performance features are
the ability to aggregate packets 
in order to reduce transmission overheads in the PHY and
MAC layers, and to use Reverse Direction (RD) which
enables stations to exchange frames without the need to contend
for the channel. We now describe these features.

Frame aggregation is a feature of the IEEE 802.11n and IEEE 802.11ac
that increases Throughput by sending
two or more consecutive data frames in a single transmission,
followed by a single
acknowledgment frame, denoted {\it Block Ack}
(BAck). Aggregation schemes benefit from amortizing
the control overhead over multiple packets. The achievable benefit
from data aggregation is often interesting, especially in light of
several factors that can impact its performance, e.g., 
link rates, collisions, error-recovery schemes, 
inter-frame spacing options, QoS guarantee etc. 
IEEE 802.11n introduces, as a pivotal part of its MAC enhancements, three
kinds of frame aggregation mechanisms:
The  Aggregate MAC Service Data Unit (A-MSDU) aggregation,
the  Aggregate MAC Protocol Data Unit (A-MPDU)
aggregation and the Two-Level aggregation
that combines both A-MSDU and A-MPDU. 
The last two schemes group several Mac Protocol Data Units (MPDU)
frames into one large frame.
IEEE 802.11ac also uses these three aggregation schemes, but enables
larger frame sizes. 

The basic idea behind the Reverse Direction (RD) feature
is a time
interval denoted Transmission Opportunity (TXOP).
A station gains a TXOP by gaining access
to the wireless channel 
and in a TXOP the station
can transmit several PHY Protocol
Data Units (PPDU) without interruption.
This station is denoted the {\it TXOP holder}.
The TXOP holder can also allocate
some of the TXOP time interval to one or more
receivers in order to allow data transmission in the 
reverse link. This is termed {\it Reverse Direction (RD)}.
For scenarios with bidirectional traffic, such as 
Transmission Control Protocol (TCP) Data segments/TCP Acks, 
RD is very attractive because
it reduces contention in the wireless
channel (no collision).

The IEEE 802.11ac
standard also defines an Automatic Repeat-Request (ARQ)
protocol that enables a transmitter to retransmit lost MPDUs
and guarantee in-order reception of MPDUs at the receiver.
This protocol is also used to improve quality of the wireless
channel.

Another feature in IEEE 802.11ac related to QoS
capabilities is the use in Access Categories (AC). There are
4 ACs: Best Effort (BE), BackGround (BK), Video (VI) and Voice (VO).
The difference between the 4 ACs is in the parameters
that control access to the channel,
namely the {\it Arbitrary Inter-Frame Space (AIFS)} 
length and the values of $CW_{min}$ and $CW_{max}$.
These vary in the various ACs and are intended to provide
priority to traffic streams with QoS requirements such as 
Video and Voice.

\subsection{Research question}

In this paper we investigate a model to transmit
TCP traffic in an infrastructure IEEE 802.11 that
optimizes the combined performance of the IEEE 802.11
MAC layer and the L4 TCP protocol using
the new features that were developed in the
latest generation of the IEEE 802.11, i.e. IEEE 802.11ac. These features
enable to use IEEE 802.11 in a completely different way than
before, as we now specify.

The issue of TCP performance over IEEE 802.11 networks
has been investigated in many papers in the past, 
e.g.~\cite{PRRSS, MKA, YC, KIA, BCG1, BCG2, BCG3, SP, NML, LCMN, KPK,
SA4, YCQ, KAMG}. 
In this past research however, it is assumed that stations
compete to get access to the channel using the contention
based CSMA/CA access method. 
Collisions are possible between stations that are involved in
different TCP connections and between the AP and stations
with which it has TCP connections due to the exchange
of TCP Data/Ack segments. Both the AP and the station(s)
try to get access to the channel simultaneously and this
results in collisions.

As far as we know, there was no development of models
to transmit TCP traffic over IEEE 802.11 using new features
of the standard.
In the model we suggest in this paper the
AP controls the TCP transmissions in the cell by configuring
the stations to use large BackOff intervals such that effectively they
never gain access to the channel and the AP enables
the stations to transmit only through time
periods delivered by the AP, the TXOP holder,
i.e. the Reverse Direction (RD)
capability. The AP communicates with one station during a TXOP
and in this paper we evaluate the performance
of such communication between the AP and the station.
Establishing policies for the communication between the
AP and several stations with which it maintains TCP
connections is the issue for
further research. 

Therefore, as a first step we assume that the AP communicates with
a single station and they have a single TCP connection between them.
The AP is the TCP transmitter, transmitting TCP Data
segments, and the station is the TCP receiver
transmitting TCP Acks.
Our performance criteria is the {\it Goodput}, defined
as the number of 
MSDUs' bits (TCP Data segments) that are
successfully transmitted and acknowledged by TCP Acks,
in the wireless channel, on average, in a second.
Such a scenario is possible, for instance in a home
environment where a Network-Attached Storage (NAS) device~\cite{L}
is attached to the AP, and a PC downloads data files
from the NAS device. It uses the aggregation and RD
capabilities of the IEEE 802.11ac MAC layer.

We use the four features of
the MAC layer of the IEEE 802.11ac mentioned above,
namely aggregation,
Reverse Direction (RD), the 
ARQ protocol 
and the four Access Categories.
Concerning aggregation, we assume the Two-Level
aggregation scheme. This scheme enables transmission of
several TCP Data segments and several TCP Acks
in a single transmission over the wireless medium.
Up to 64 MPDUs can be transmitted in a single
transmission and every MPDU can contain
several MSDUs. We measure the influence of
aggregation on the Goodput.

Notice that in a TCP connection over IEEE 802.11ac both
sides of the connection compete for the wireless channel - one for
transmitting TCP Data segments and the other for
transmitting TCP Acks. This competition can result
in collisions and reduced Goodput.
We examine two operation modes for the transmission of TCP traffic over
the wireless medium.
In one operation mode, using RD,
the TCP transmitter allocates a 
TXOP when
it acquires the wireless medium, and enables
the TCP receiver to transmit TCP Acks during the TXOP
without collisions. 
In the second operation mode, for
comparison purposes, which we denote by {\it No-RD}, 
the traditional CSMA/CA random access
MAC is used.
The TCP transmitter and the TCP receiver contend for the
wireless medium in every transmission attempt.
The operation mode using RD is more complicated
than the contention based one, and we want to check if, and
to what degree, using RD improves the Goodput
of the {\it No-RD} operation mode.

In addition to all the above we assume the ARQ protocol
of the IEEE 802.11ac standard at the MAC layer.
This protocol guarantees an in-order delivery
of MPDUs between communicating entities.
However, due to its Transmission Window, 
the ARQ protocol can sometimes limit the number
of MPDUs transmitted in each transmission, i.e.
this protocol can limit the amount of aggregation.

Finally, we check the influence of the values
of the access parameters in the four ACs
on the
Goodput, namely the Arbitrary Inter Frame Space (AIFS),
Contention Window min.,
$CW_{min}$, and Contention Window max. $CW_{max}$.

We assume that the AP and the station
are the end points of the TCP connection.
Following e.g.~\cite{MKA,BCG1,BCG2,KAMG}
it is quite common to consider short Round Trip Times (RTT)
in this kind of high speed networks such that
no retransmission timeouts occur. Notice also
that due to the MAC ARQ protocol, the L4 TCP protocol
always receives TCP Data segments in order. Therefore,
the TCP congestion window increases up to the
TCP receiver advertised window. We assume that the
TCP receiver window is large enough such that the
TCP Transmitter Transmission window can always provide
as many MSDUs to transmit
as the MAC layer enables. We assume the above
following the observation
that aggregation is useful in scenario where the offered
load on the channel is high. We therefore do not consider
the TCP Transmission Window and our goal
is to find the maximum possible Goodput that the
wireless channel enables to a single TCP connection,
where the TCP itself does not impose any limitations
on the offered load, i.e. on the rate that MSDUs are
given for transmission to the
MAC layer of the IEEE 802.11ac.

Following the above we also do not consider a particular
flavor of TCP, e.g. TCP NewReno, Westwood, Cubic~\cite{HFGN,MCGSW,HRX}
if to mention only a few. All the TCP flavors differ in the
way they handle the TCP congestion window but in this
paper, as mentioned, we assume that the TCP Transmission Window
is limited only by the TCP receiver advertised window.

Regarding the wireless channel quality we first
assume an error-free channel, i.e. the
Bit Error Rate (BER) equals 0.
Then we assume another
three BERs : $10^{-5}, 10^{-6}$
and $10^{-7}$.
The scenario of a single TCP connection 
with various
BER values
is possible for instance in the mentioned home environment where
a Network-Attached Storage (NAS)
device~\cite{L} is attached to the AP,
and a PC, which is a client in the IEEE 802.11
system, is located close to the NAS and
downloads data files from the NAS device.
The various BERs are a function
of the channel conditions between the client (e.g. PC) and
the AP. If they are stable and have a low path loss channel
between them.
the BER is very low. However,
if the PC is located in the basement for instance, the BER
can be larger.

An additional feature that we use was introduced in~\cite{SA2}. In~\cite{SA2}
a repetition scheme is introduced, in which several MPDUs in a 
single transmission are transmitted several times. This feature
improves the achieved Goodput in large BERs, as will later become
clear. 

\subsection{Our results}

We show that for an error-free channel, i.e. BER$=$0, using RD 
improves the Goodput over not using RD by $20\%$. Moreover, using
TXOPs of about $20 \mu s$ are sufficient to achieve that
improvement, and this outcome has an impact on the delay at the
TCP protocol from the time the TCP transmitter transmits TCP Data
segments until it receives the corresponding TCP Acks.

For error-prone channels we show that using RD improves the Goodput
in almost $50\%$ and when also using the Repetition scheme of~\cite{SA2} the
improvement can even reach $60\%$. TXOPs of about $4 \mu s$ are
sufficient to achieve these Goodput improvements.

\subsection{Previous works}

From the point of view of Transport protocols,
the performance of the IEEE 802.11 protocol has been
investigated in two models : UDP-like traffic
and TCP traffic, i.e. when there is
bi-directional traffic that can result
in collisions. By UDP-like traffic we mean that
the Data receiver does not transmit an Ack at the
Transport layer, nor, in terms of IEEE 802.11,
does it generate an MSDU for transmission.
In TCP traffic, the receiver of TCP Data segments
generates an MSDU which contains a TCP Ack,
and depends on the channel for its transmission.

Regarding UDP-like traffic,
the performance of IEEE 802.11 (taking into account
the aggregation schemes)
has been investigated in dozens of papers over the years.
For example,
in~\cite{SC,LW,GK,SNCSKJ,KHS,C1,WW,SS,Z,DAM,SOSH,KKS,KMLPC}
the Throughput and Delay performance 
of the A-MSDU, A-MPDU and Two-Level aggregation schemes are
investigated. Several papers assume an error-free channel
with no collisions,
several papers
assume an error-prone channel and others also
assume collisions. In~\cite{P,OKACHN,CAHNOKR,BBSVO, SA}
the performance of 802.11ac is investigated. Papers~\cite{OKACHN,SA}
consider the performance of the aggregation schemes in 802.11ac
and compare the performance of 802.11ac to that of 802.11n. 

Another set of papers~\cite{CS,SNMB,TQDDYT,SNZ,MGCR,SA1}
deals with QoS together with the aggregation schemes. 
In particular, in~\cite{SA1} the use of the ARQ protocol
of the IEEE 802.11 standard~\cite{IEEEBase1}, together with the 
aggregation schemes, is investigated in relation to QoS guarantee.

Concerning TCP traffic, we can specify
a first set of papers that deal with TCP's Throughput,
Delay and Fairness performance over
legacy IEEE 802.11/a/b/g networks. There are dozens
of such papers, such as~\cite{PRRSS,MKA,YC,KIA,BCG2,BCG3,SP} to mention
only a few. None of the papers from this set
consider Access Categories or aggregation schemes
that were introduced in later versions of the standard, i.e.
IEEE 802.11e and IEEE 802.11n respectively. 

As the IEEE 802.11e was introduced, many papers
appeared concerning this standard and the performance of
TCP. In IEEE 802.11e the Access Categories are
defined, enabling change to the fix values of
the DIFS ( now called AIFS ) and $CW_{min}$ of the
previous versions of the standard.
Also introduced is the TXOP
time interval that enables the AP/stations to transmit
several frames in a single transmission opportunity.
Such frames are acknowledged in the MAC layer, all together,
by a new defined frame; the {\it Block-Ack} frame.
Papers regarding TCP investigate the use of the
above changes in improving TCP performance~\cite{NML,LCMN}.
None of the papers concerning IEEE 802.11e and TCP
deal with ACs and aggregation schemes, as does our paper,
since aggregation schemes were only introduced in a
later version of the standard, namely IEEE 802.11n.

In relation to IEEE 802.11n/ac where aggregation
is introduced, we are aware of only three research papers
that handle the Throughput performance of TCP
in the various aggregation schemes\cite{GK,KPK,SA4}.
In~\cite{GK} the authors also assume the model
of the AP and a single station that maintain a TCP connection. The 
paper considers the A-MSDU and A-MPDU aggregation 
schemes only, and does not
consider the Two-Level aggregation scheme, the RD and the
various ACs. In the analysis the authors
assume a TCP Transmission window of one TCP Data
segment. On the other hand, in this paper we also
handle the Two-Level aggregation scheme, the various
ACs, the RD and a TCP Transmission window larger than one
Data segment, which complicates the analysis.

In~\cite{KPK} 
it is argued that aggregation increases the discrepancy
among upload TCP connections. The model is an AP with several
stations that initiate TCP upload connections. The A-MPDU aggregation
is considered and there is no a reference to Two-Level
aggregation, to RD and to
the standard ACs. 
The authors suggest an algorithm
to reduce the discrepancy among TCP connections.
Our paper deals with another model: we explore the influence
of aggregation on the Goodput of a {\it single} TCP connection,
e.g. in a home environment, consider Two-Level
aggregation, RD and check the performance of the 4 ACs defined
in IEEE 802.11 . 

In~\cite{SA4} the performance of a single TCP connection
is evaluated using all three aggregation schemes and
four standard ACs. However, only an error-free channel is
considered and there is no reference to RD, i.e. there can
be collisions between the two parties of the TCP connection.
The current paper is a next step to the research in~\cite{SA4}
in the sense that it also considers an error-prone channel and 
explores the elimination of collisions by RD.

Regarding RD, there are several papers such 
as~\cite{GSV,MNSA,MNSAM,PFGZG} that deal with
RD's Goodput performance, also in relation to TCP.
However, these papers do not consider aggregation, ACs and the
IEEE 802.11ac ARQ protocol all together. 

Finally, none of the papers mentioned in this literature survey
consider the Repetition scheme of~\cite{SA2} and
its influence on the Goodput performance.

The rest of the paper is organized as follows: In Section 2
we describe in detail the features of the IEEE 802.11ac
that we use in this paper. In Section 3 we describe
the model we suggest for TCP transmission over IEEE 802.11  
using RD. In Sections 4 and 5 we compute
the Goodput performance of the error-free and error-prone
channels respectively. Section 6 concludes the paper
and in the Appendix we present a Markov chain model for the scenario
in which there is no use in RD and both the AP and the station
contend for the channel in every transmission attempt.

\section{Network Model}

\subsection{Aggregation schemes}

Three aggregation schemes are defined 
in IEEE 802.11n/ac: Aggregate
MAC Service Data Unit (A-MSDU), Aggregate MAC Protocol Data
Unit (A-MPDU) and Two-Level aggregation, which combines the former two.

\begin{figure}
\vskip 12cm
\includegraphics{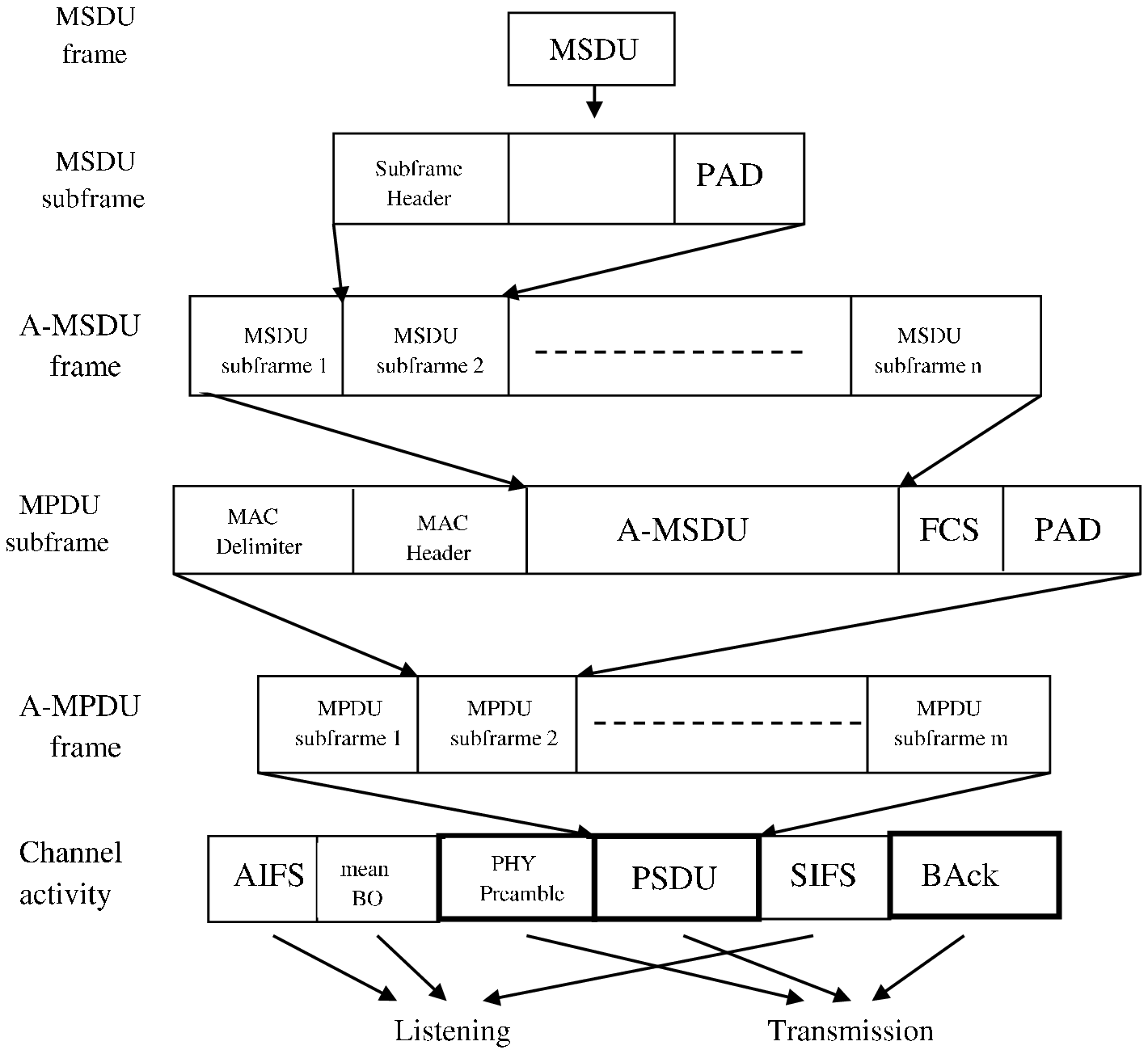}
\caption{The Two-Level aggregation process.}
\label{fig:twole}
\end{figure}

In A-MPDU aggregation several MPDUs are combined together
into a single PHY Service Data Unit (PSDU) denoted A-MPDU frame, and are
transmitted in one PHY Protocol Data Unit (PPDU), thus
saving PHY overhead.  
The Two-Level aggregation scheme is shown in Figure~\ref{fig:twole}.
In this aggregation scheme several MPDUs are again inserted for
transmission into one A-MPDU frame.
However, an MPDU can contain
several MSDUs.
Every MSDU is preceded by a SubFrame Header of
14 bytes and every MSDU, with its SubFrame Header,
is rounded by a PAD to a size that is an integral
multiple of 4 bytes.
Every MPDU is preceded
by a MAC Delimiter of 4 octet and is rounded by a PAD
with its delimiter, to a length that is an integral
multiple of 4 octets. 
Such MPDUs are denoted A-MSDU frames.
The Two-Level aggregation scheme achieves a better
ratio than the other aggregation
schemes between the amount of Data octets transmitted to the
PHY and MAC layers' overhead. 

In 802.11ac the maximum A-MPDU's size is
1048575 octets and the MPDUs' maximum
length is 11454 octets.
The maximum number of MPDUs in an A-MPDU frame is 64.

Assume an A-MPDU frame that contains $K$ MPDUs and $Y$ MSDUs.
Let $L$ be the length of an MSDU in bytes. Recall that according
to the IEEE 802.11ac standard~\cite{IEEEBase1, IEEEac},
every MSDU within an MPDU frame is preceded 
by a  SubFrame Header of 14 bytes, and with this field
it is rounded to an integral multiple
of 4 bytes. The received size is $L^{'}$ such that
$L^{'}= 4 \cdot \ceil{\frac{L+14}{4}}$.
Recall also that every MPDU within a PSDU is
preceded by a MAC Delimiter. To compose an MPDU
one also adds the MAC Header and 
Frame Control Sequence (FCS) fields to the MSDUs of the MPDU.
Thus, the 
length $Len$ of the PSDU
in bytes is  $Len =(K \cdot H) + (Y \cdot L^{'})$ 
where $H= 4 \cdot
\ceil{\frac{MacDelimiter + MacHeader + FCS}{4}}$.
{\it MacDelimiter, MacHeader}
and {\it FCS}
denote the sizes in bytes of the MAC delimiter, MAC Header and
FCS fields respectively. 

The receiver of an A-MPDU frame acknowledges its reception
by a Block Ack (BAck) control frame. In BAck the receiver
separately acknowledges the reception of every MPDU in the
received A-MPDU frame.

Let $O_1 = AIFS + Preamble + SIFS + BAck$ and let $BackOff$ be
the BackOff interval that a station uses in a given transmission.
The transmission time without 
collisions of the above A-MPDU is~\cite{IEEEBase1}:

\begin{equation}
O_1 + BackOff + TSym \cdot \ceil{\frac{8 \cdot Len+22}{BitsPerSymbol \cdot R}}.
\label{equ:twole1}
\end{equation}

The additional 22 bits are due
to the SERVICE ( 16 bits ) and TAIL ( 6 bits ) fields
that are added to every transmission by the PHY layer 
Conv. Protocol~\cite{IEEEBase1}.

In Eq.~\ref{equ:twole1} we assume the OFDM PHY layer.
$Tsym$ is the duration of one
Transmission Symbol in OFDM, and it is $4\mu s$. $BitsPerSymbol$ equals
4 in OFDM and $R$ is the PHY rate in Mbps. Any transmission
in OFDM must be of an integral number of Symbols.

\subsection{The Error model}

We assume that the process
of frame loss in a wireless fading channel
can be modeled with a good approximation by a low order Markovian 
chain, such as the two state
Gilbert model~\cite{L1,ZRM}.

In this model the state diagram is composed of
two states, "Good" and "Bad", meaning successful or unsuccessful
reception of every bit arriving at the receiver, respectively.
{\it Bit-Error-Rate} (BER) is the 
probability of moving from the Good state to the
Bad state.
($1-BER$) is the probability of staying at the Good state.
According to the above model, the success probability of a frame
of length $B$ bits is $(1-BER)^B$ and the failure probability
$p$ is given by Eq.~\ref{equ:failureprobbasic}:

\begin{equation}
p=1-(1-BER)^{B}
\label{equ:failureprobbasic}
\end{equation}

By the above model
one can see that as the frame length $B$ increases, so does the
failure probability. Thus, in every aggregation scheme,
increasing the aggregation amount increases the frame's size
as well as the transmission delay of the frame. The 
failure probability can sometimes also
increase. 

We would like to mention that there are other models to
represent the quality of the indoor wireless channel,
e.g. the one in~\cite{GAGM}. This model shows burstiness in the
channel quality. In this paper however, we assume that the
communicating stations use Link Adaptation by which
they keep the effective SNR stable and in such a scenario the
BER is stable.

\subsection{IEEE 802.11ac ARQ protocol}

We give only a brief description of the IEEE 802.11ac
ARQ protocol.
A more detailed description can be found in ~\cite{SA1}
and in sections 9.21.7.3 - 9.21.7.9 in~\cite{IEEEBase1}.

Consider the transmission of a series  of
MPDUs from one entity to another
in IEEE 802.11ac . MPDUs are numbered, and the recipient
signals the transmitter which MPDUs arrived successfully and
which in error. Failed MPDUs are retransmitted by the transmitter.
The number of retransmissions of an MPDU is limited.

The transmitter maintains a Transmission Window (TW) over
the sequence numbers of the MPDUs. We denote this
transmission window by MAC TW, to distinguish it
from transmission windows of higher levels' protocols,
such as the one of TCP. Only MPDUs within the MAC TW
are allowed for transmission to the recipient. The
maximum size of the TW is 64 consecutive sequence numbers because
the recipient can acknowledge at most 64 MPDUs in one
BAck control frame~\cite{IEEEBase1}.

Let X be the smallest sequence number in the MAC TW and X+63 be
the largest. As long there is no acknowledgment
from the recipient that MPDU X arrived successfully, the
MAC TW does not change. When an acknowledgment for MPDU X
arrives, the MAC TW moves one position (number) along the sequence
numbers' space : X is taken out and X+64 is inserted into the MAC TW.

Let $K$ be the maximum number of MPDUs that can
be transmitted in one PPDU in Two-Level aggregation.
Assume that MPDU X has been transmitted several times
with no success. In this case the MAC TW is unchanged and
it is possible that only $M$ MPDUs within the MAC TW,
$M<K$ are unacknowledged by the recipient.
In such a case only $M$ MPDUs are transmitted and
the MAC TW limits the number of transmitted MPDUs.
As $K$ is larger the probability for such a scenario
is larger.

\subsection{Timing}

We assume the following values for the time intervals
used in IEEE 802.11ac and we assume that the reader
is familiar with the basic access scheme of IEEE 802.11ac networks. For
the OFDM PHY layer SlotTime$=9 \mu s$ and SIFS$=16 \mu s$.
The BAck and Ack 
frames are 32 and 14 bytes
long respectively.
Their transmission
times, denoted $BAckTime$ and $AckTime$ respectively are $32 \mu s$
and $28 \mu s$ respectively, using
the Basic PHY Rate of 24Mbps. These
times include the PHY preamble preceding the transmissions
of these frames.
If the PHY rate $R$ used for data frame transmissions
is lower than 24Mbps, then $R$ is also 
used for the BAck and Ack transmissions.
However, in this paper we
assume a PHY rate of 1299.9 Mbps assuming working point
MCS9 with 3 spatial streams and an 80MHz channel.
With 3 spatial streams the PHY Preamble is $48 \mu s$~\cite{IEEEBase1}.

\subsubsection{Successful transmissions}

\begin{figure}
\vskip 4cm
\includegraphics{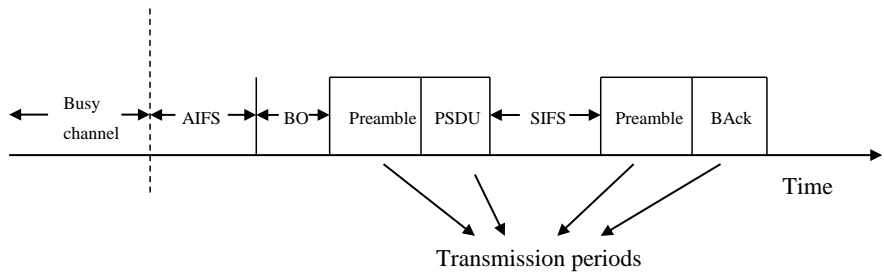}
\caption{The activity on the channel in the case of a successful transmission.}
\label{fig:success}
\end{figure}

In Figure~\ref{fig:success} we show the activity on the
channel where a successful transmission occurs, i.e. without
collisions. 
In this case, after a station senses
an idle channel for a duration equal to its
AIFS and BackOff
intervals, it transmits the data frame. After a SIFS and 
a PHY Preamble the receiver
acknowledges reception.  In the case of
Two-Level aggregation the BAck frame is used.

\subsubsection{Collision events}

\begin{figure}
\vskip 5cm
\includegraphics{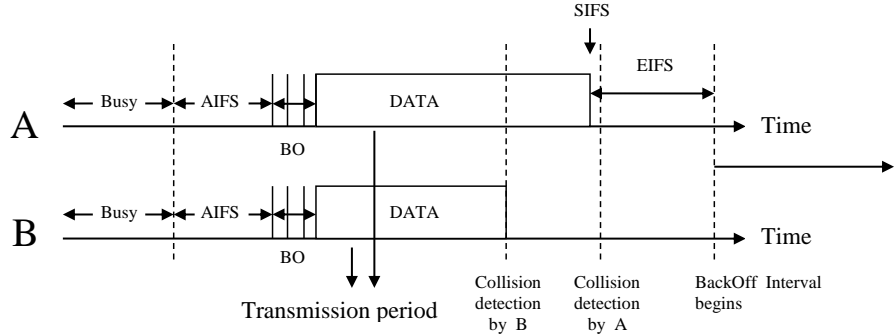}
\caption{The activity on the channel in the case of a collision.}
\label{fig:collision1}
\end{figure}

In Figure~\ref{fig:collision1} we show
the activity on the channel
in the event of collisions.
We show two stations, A and B . After the channel is clear,
and assuming that the Network Allocation Vector (NAV)
at both stations is equal to 0,
both stations wait the AIFS interval. If
their BackOff intervals are
equal, both stations
begin together to transmit their data frames when
the BackOff intervals terminate. If the data frame of B
is shorter than that of A, then when B terminates its transmission
it detects a carrier on the channel; that of A's transmission.
Thus, it recognizes that they have collided. When A terminates its 
transmission, it waits the SIFS interval, recognizes that it has not
received an acknowledgment and so detects the collision. 
Both A and B now
wait the interval Extended Inter Frame Space (EIFS)
after the transmission of A terminates.
The channel then becomes clear and the BAckOff
intervals at the stations start again. EIFS is the interval that stations
wait in IEEE 802.11ac after a collision is detected~\cite{IEEEBase1}.

\subsection{Access Categories}

As mentioned, we consider the four ACs
defined in the IEEE 802.11ac standard, i.e.
BE, BK, VO and VI. The ACs defer in the
values of the parameters
that control the access to the channel, namely
AIFS, $CW_{min}$ and $CW_{max}$.
For every AC the value of AIFS, denoted AIFS[AC], equals to 
$SIFS + AIFSN[AC] \cdot SlotTime$. 
The various Access Category Numbers (AIFSN[AC]), $CW_{min}$ and $CW_{max}$
appear in Table~\ref{tab:model1}. The values are taken
from the WiFi Alliance (WFA) publications~\cite{WFA}.
The EIFS used
in every AC, denoted $EIFS[AC]$, equals to $SIFS + AckTime + AIFS[AC]$.
For the computation of EIFS, it is assumed that the Ack
frame is transmitted in the smallest basic PHY rate of 6Mbps, i.e. 
the $AckTime$ for the
computation of the EIFS is $44 \mu s$.
The value of AIFS[AC] and EIFS for every AC also appear
in Table~\ref{tab:model1}.

\begin{table}
\caption{\label{tab:model1}{The values of $CW_{min}$,
$CW_{max}$, AIFS number, AIFS and EIFS in the 
four Access Categories of IEEE 802.11ac for a station (Access Point)}}
\vspace{3 mm}
\center
\begin{tabular}{|c|c|c|c|c|}  \hline
 & BK  &  BE  & VI  &  VO \\ \hline
$CW_{min}$  & 16(16)  &  16(16) &  8(8) & 4(4) \\ \hline
$CW_{max}$  & 1024(1024)  & 1024(64)  & 16(16) & 8(8) \\ \hline
AIFSN  & 7(7)  &  3(3) & 2(1) & 2(1) \\ \hline
AIFS[AC]($\mu s$) & 79(79) & 43(43) & 34(25) & 34(25) \\ \hline
EIFS[AC]($\mu s$) & 139(139) & 103(103) & 94(83) & 94(83) \\ \hline
\end{tabular}  
\end{table}

\section{TCP traffic model over IEEE 802.11}

\begin{figure}
\vskip 15cm
\includegraphics{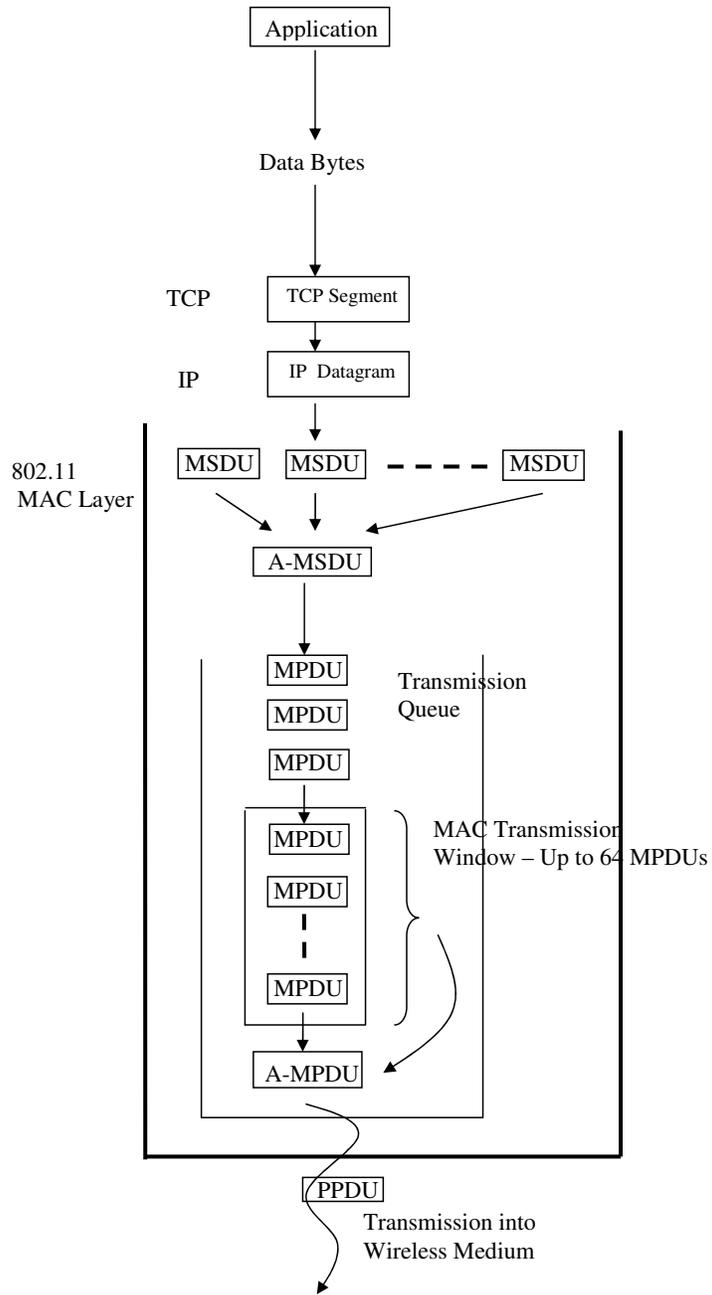}
\caption{The Traffic model.}
\label{fig:traffic}
\end{figure}

In this section we describe our model for the transmission
of TCP traffic over IEEE 802.11 .
In Fig.~\ref{fig:traffic} we show
the Traffic flow considered.

We assume an application with a massive
Data stream, for example a Video stream that 
uses TCP, and its Data bytes are mapped
into equal length TCP segments. A TCP segment
is mapped into an IP Datagram. IP Datagrames are given
to the MAC layer of the IEEE 802.11ac as MAC Service Data Units (MSDU)
and these are packed into MPDUs (A-MSDUs). MPDUs are kept in a 
Transmission Queue and
are transmitted using Two-Level aggregation.
Recall that we assume a saturation scenario in which the TCP always has
an unlimited number of Data segments to transmit.
Also notice the MAC TW mentioned in Section 2.3 .

\subsection{Operation modes for TCP Usage of the channel}

We consider 2 operation modes for the transmission of TCP Data/Ack
segments over the channel.

\subsubsection{Operation mode 1 - No-RD, Competition}

Both the TCP transmitter and the TCP receiver
contend for the channel in every transmission attempt,
i.e. when the TCP receiver has TCP Acks to transmit,
it contends for the channel with the TCP transmitter
in every transmission .
Both stations use the Two-Level aggregation. 


\subsubsection{Operation mode 2 - Reverse Direction}

Reverse Direction is a mechanism in which the
owner of a Transmission Opportunity (TXOP)
can enable its receiver to transmit
back during the TXOP, so that the receiver does not need
to contend for the channel. This is particularly efficient for
a bi-directional traffic such as TCP Data segments and TCP Acks.

We examine an operation mode in which the TCP transmitter (AP) transmits
A-MPDU frames containing MPDUs of TCP Data segments
to the TCP receiver (station), and enables
the TCP receiver to answer with an A-MPDU frame containing
MPDUs frames of
TCP Acks. Both stations use the Two-Level aggregation.

We assume the following scenario to use
RD, as
is illustrated in Fig.~\ref{fig:rd3}:

\begin{figure}
\vskip 9cm
\includegraphics{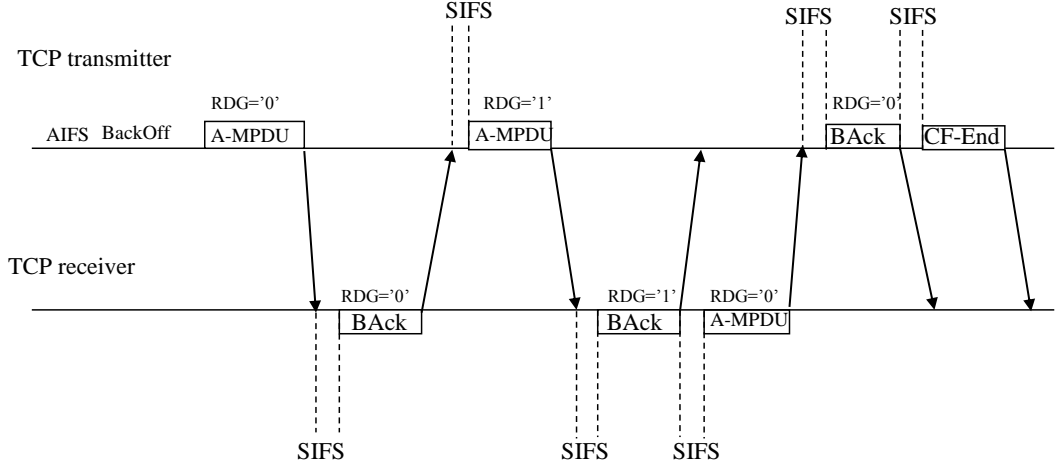}
\caption{The operation mode that uses Reverse Direction.}
\label{fig:rd3}
\end{figure}

After waiting AIFS and BackOff the TCP transmitter (AP)
transmits $n$ A-MPDU frames in a row in the TXOP. 
In Figure~\ref{fig:rd3} we assume $n=2$. The TCP
receiver (station) responds to every transmission
by a BAck frame. In its last A-MPDU frame the
TCP transmitter sets the RDG bit~\cite{IEEEBase1}, enabling the TCP
receiver to respond with an A-MPDU frame.
The TCP transmitter then responds with a BAck frame
and terminates the TXOP with the CF-End frame~\cite{IEEEBase1}.

We assume that there are no collisions on the channel
after the end of a TXOP because the TCP receiver
is configured in a way that prevents collisions. For example, the
TCP receiver is configured 
to choose its BackOff interval
from a very large contention interval, other than
the default ones in Table~\ref{tab:model1}.  Thus, the
TCP transmitter always wins the channel without collisions.
The transmissions on the channel are composed 
of TXOPs that repeat themselves one after the other.
We denote by $RD(n)$ the case where the TCP transmitter
transmits $n$ A-MPDU frames in the TXOP.

\section{Error-Free Channel Results}

In this section we assume an 
error-free channel, i.e. BER$=$0, and in this
case the operation mode using
Reverse Direction (RD)
is as follows: Every transmission of the TCP transmitter
contains $K_D$ MPDUs, $1 \le K_D \le 64$. Assuming
TCP Data segments of $L_{DATA}=1480$ bytes, the resulting
IP Datagrams are of 1500 bytes ( 20 bytes of IP Header are added ) 
and together with the SubHeader field and rounding
to an integral multiple of 4 bytes, every MSDU is
of $L^{'}_{DATA}=1516$ bytes. Due to the limit
of 11454 bytes on the MPDU size, 7 such MSDUs are possible in one MPDU.
The total number of MSDUs transmitted by the
TCP transmitter in one TXOP is therefore $n \cdot K_D \cdot 7$.

The TCP receiver transmits TCP Acks. Every TCP Ack
is of $L_{Ack}=48$ bytes ( 20 bytes of TCP Header + 20 bytes of IP header +
8 bytes of LLC SNAP ). Adding 14 bytes of the SubHeader
field and rounding to an integral multiple of 4 bytes, every
MSDU of the TCP receiver is $L^{'}_{Ack}=64$ bytes, and every
MPDU, again due to the size limit
of 11454 bytes, can contain 178 MSDUs.
Every transmission
of the TCP receiver can contain
up to 64 MPDUs.

The receiver can transmit up to $64 \cdot 178$ TCP
Acks (MSDUs) in a single transmission. Therefore, the
number $n$ of transmissions of the TCP transmitter in a TXOP
should be limited by the following inequality : $ n \cdot K_D \cdot 7
\le 64 \cdot 178$.
Using larger $n$'s will not increase the Goodput.

Let $AIFS, BO, Preamble$ and $SIFS$ denote the length,
in $\mu s$, of the AIFS, BackOff, PHY Preamble and SIFS
time intervals, and {\it BAck, CF-End} denote, in $\mu s$,
the transmission times of the BAck and CF-End control frames respectively.
Let $H= MacDelimiter + MacHeader + FCS$ be the total length
of the MAC Delimiter, MAC Header and FCS fields 
of an MPDU in bytes respectively.
We assume that the MAC Header is of 28 bytes and the FCS is 4 bytes.
Therefore, $H=36$ bytes.

Since there are no collisions when using RD, holds
$BO = \frac{(CWmin-1)}{2} \cdot SlotTime$,
where we refer to the $CW_{min}$ of the AP. See Table~\ref{tab:model1}.
We now define $C$ to be {\it C=AIFS+BO+SIFS+CF-End+Preamble}. 
The last $Preamble$ in $C$ is the one preceding 
the transmission of the station.

Let $T(AP)$ and $T(STA)$ be the transmission
times of the AP and the station's A-MPDU frames respectively.
$T(AP)$ is given by the following ( the details of how Eqs. 3-5 
are derived can be found in~\cite{SA}):

\begin{equation}
T(AP) =  4 \cdot 
\ceil{\frac{K_D \cdot (L^{'}_{DATA} \cdot 7 + H) \cdot 8 + 22 }{4 \cdot R}}
\end{equation}

\noindent
and $T(STA)$ is as follows:

\begin{equation}
T(STA) = 4 \cdot 
\ceil{\frac{(n \cdot K_D \cdot 7 \cdot L^{'}_{Ack} + K_A \cdot H) \cdot 8 + 22}
{4 \cdot R}}
\end{equation}

\noindent
where $K_A$ is the number of MPDUs in the station's A-MPDU frame
and $ K_A = \ceil{\frac{n \cdot K_D \cdot 7}{178}}$.

\noindent
The length $cycle$ of a TXOP is therefore given by

\begin{equation}
cycle = C + n(Preamble+T(AP)+SIFS+BAck+SIFS) + T(STA) + SIFS + BAck
\end{equation}

\noindent
and the Goodput of the system is

\begin{equation}
Goodput = \frac{n \cdot K_D \cdot 7 \cdot (L_{DATA} \cdot 8)}{cycle}
\end{equation}

\noindent
Neglecting the rounding of $T(AP), T(STA)$ and $K_A$
the Goodput can be written as :

\tiny

\begin{eqnarray}
\label{equ:goodput}
Goodput &= &
\frac{n \cdot K_D \cdot 7 \cdot (L_{DATA} \cdot 8)}
{C+SIFS+BAck
+n \cdot (Preamble+2 \cdot SIFS + BAck)
+\frac{(n \cdot K_D(L^{'}_{DATA} \cdot 7 + 
H) \cdot 8 + 22)}{R} + 
\frac{(n \cdot K_D \cdot 7 \cdot L^{'}_{Ack}+K_A \cdot H) \cdot 8 + 22}{R}
} \\ \nonumber
& = &
\frac{L_{DATA} \cdot 8}
{\frac{C+SIFS+BAck}{n \cdot K_D \cdot 7} + 
\frac{(Preamble+2 \cdot SIFS +BAck)}{K_D \cdot 7} +
\frac{8}{7R}(L^{'}_{DATA} \cdot 7 + H) + 
\frac{22 \cdot 2}{n \cdot K_D \cdot 7 \cdot R} + 
\frac{8 \cdot L^{'}_{Ack}}{R}+
\frac{8 \cdot H}{178 \cdot R}}
\end{eqnarray}

\normalsize

\noindent
One can see that as $n$ increases and/or $K_D$ increases, so does
the Goodput.
Notice that since BER$=$0, the MAC ARQ protocol does not impose
any limitation on the number $K_D$ of MPDUs that are
transmitted by the TCP transmitter in every transmission, as
long as $K_D \le 64$. Also, it is most efficient to
contain 7 MSDUs in every MPDU because this choice best
amortizes the PHY/MAC overheads over the MSDUs.

\begin{figure}
\vskip 16cm
\includegraphics{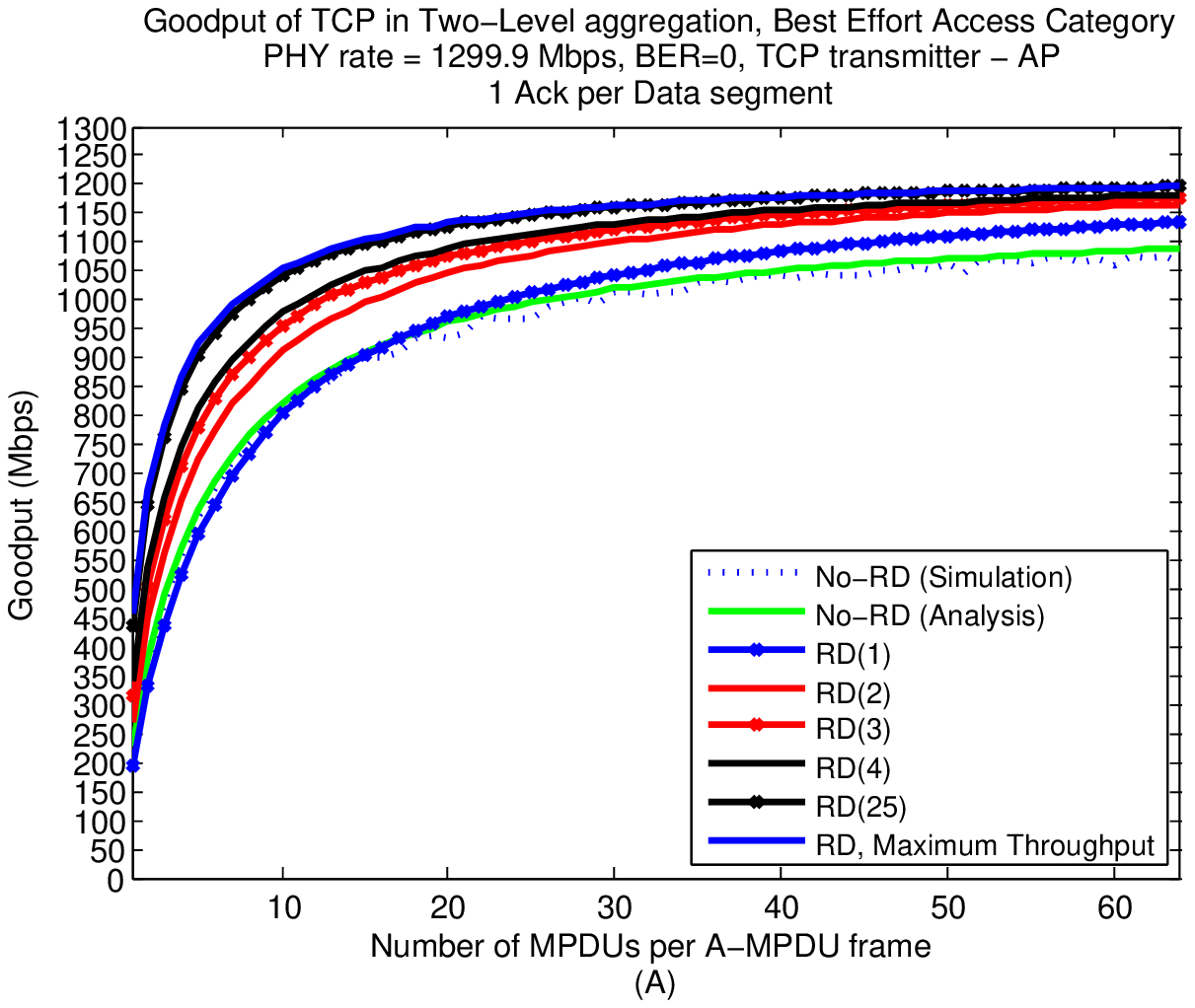}
\includegraphics{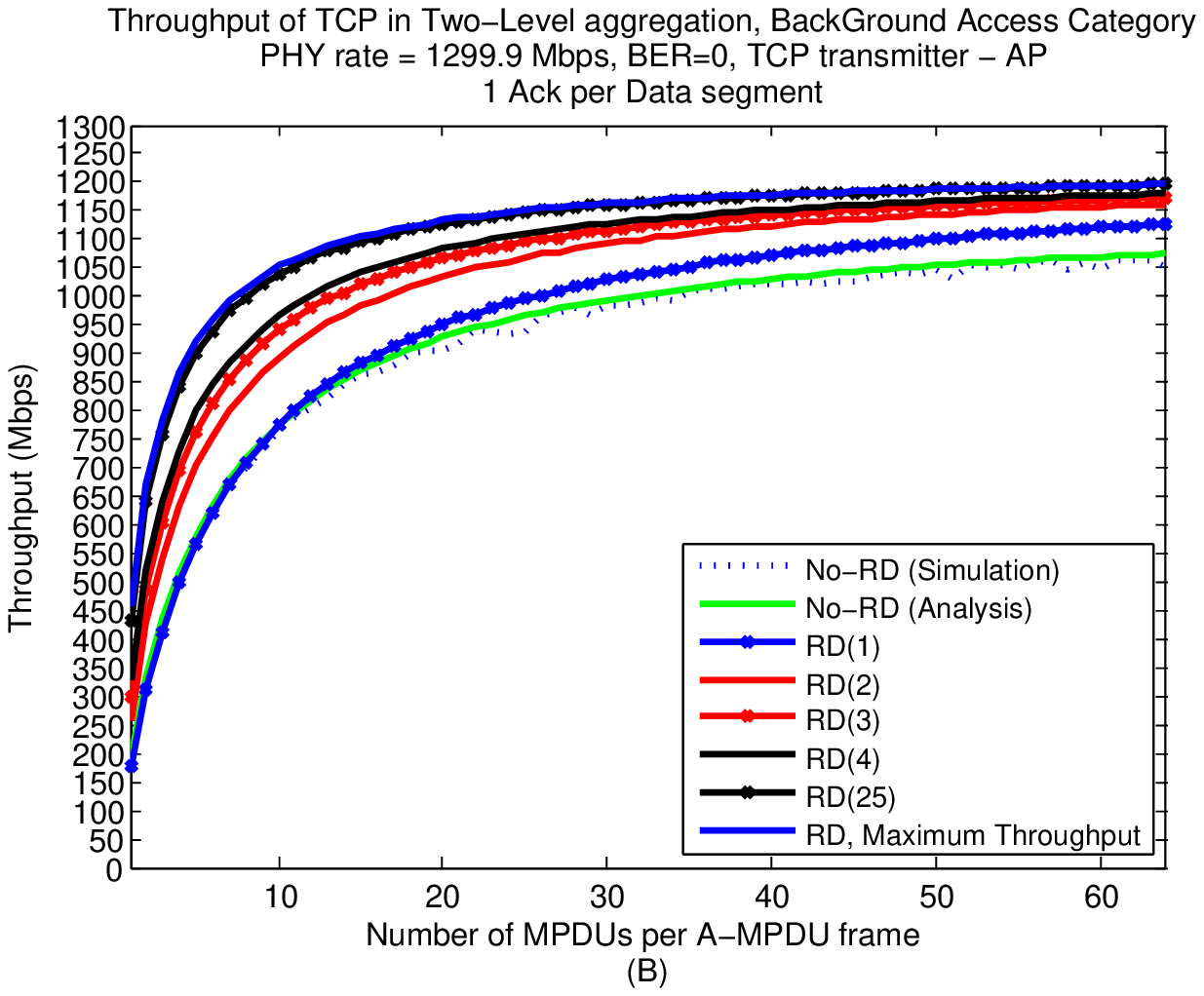}
\includegraphics{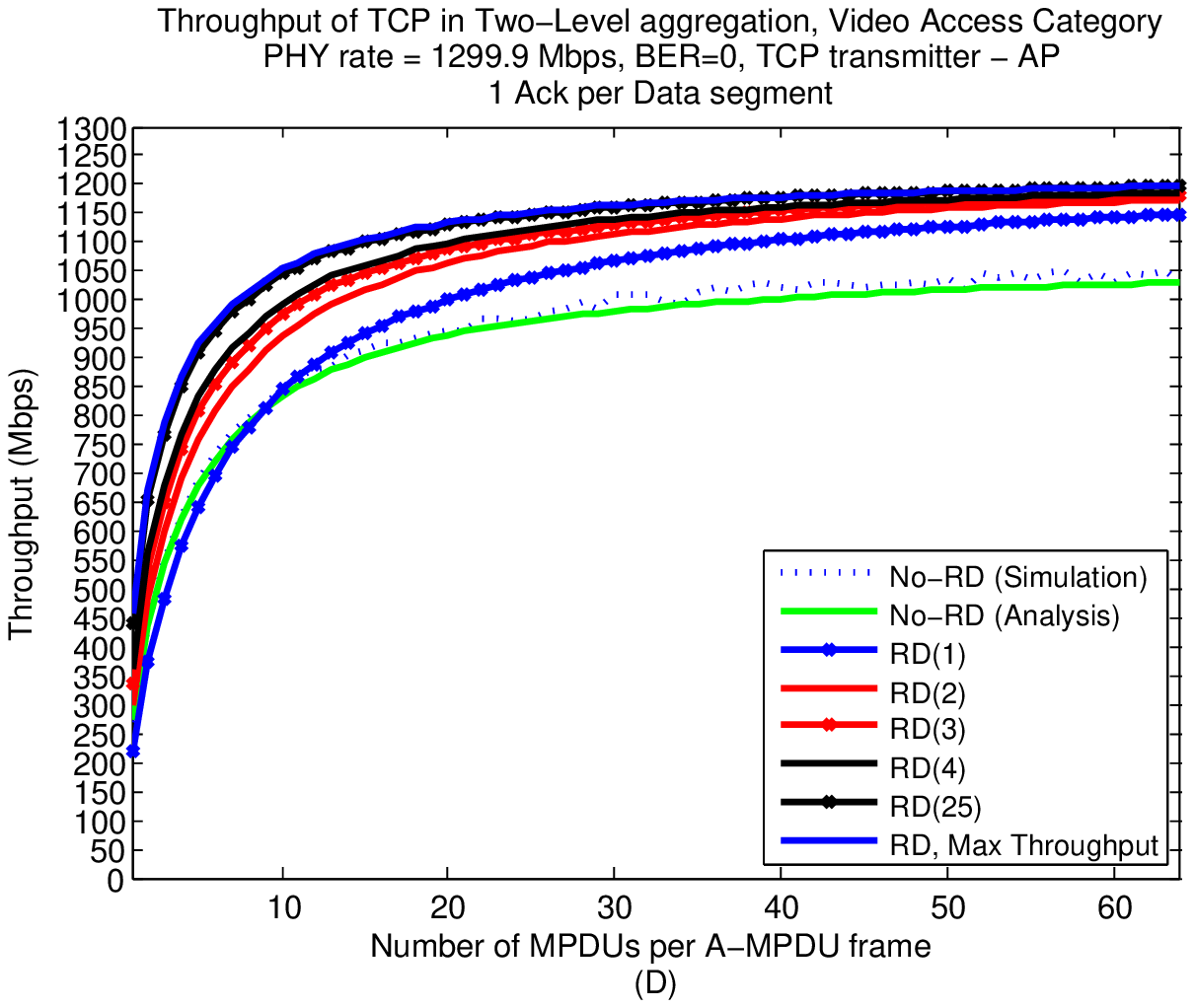}
\includegraphics{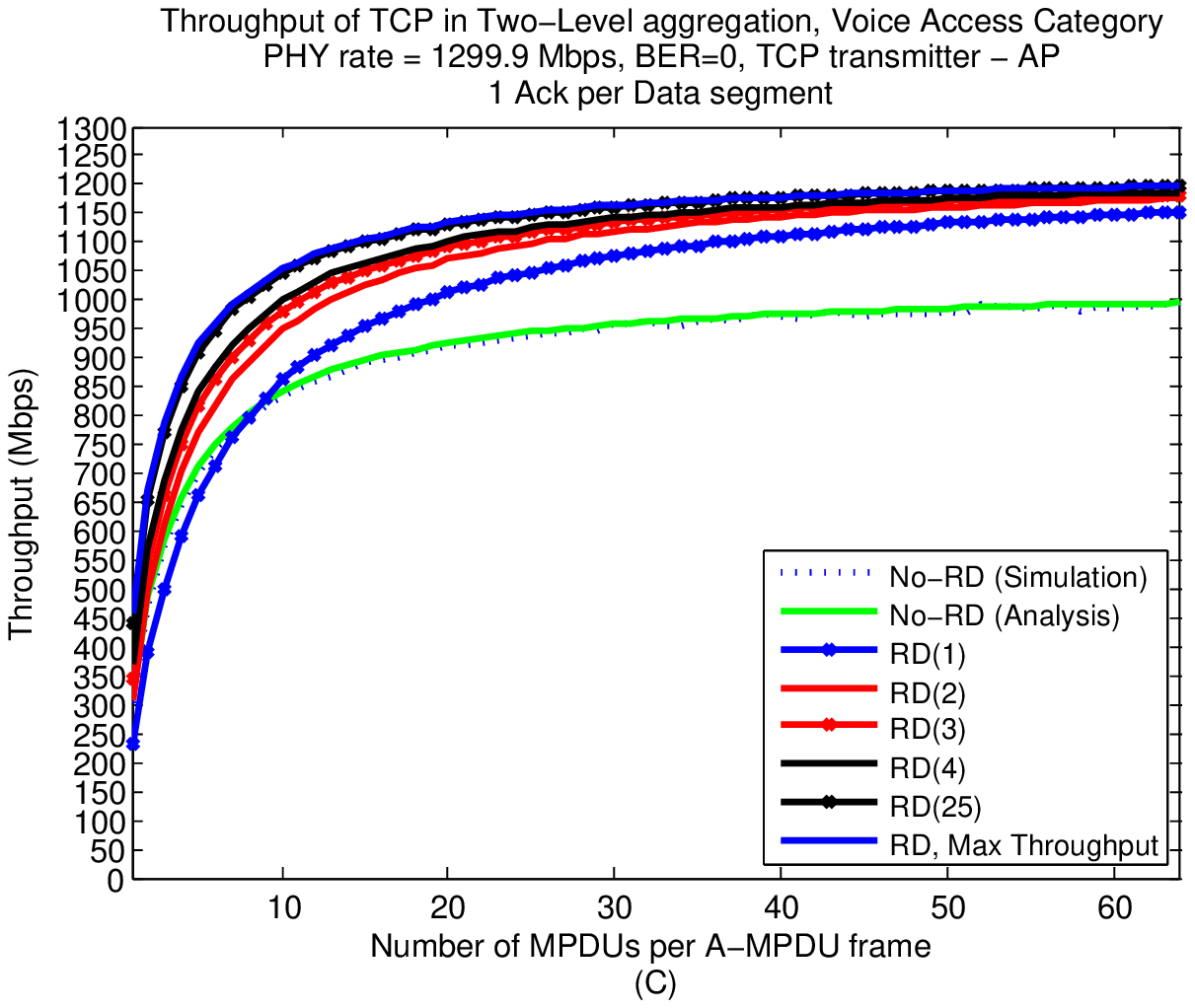}
\caption{The Goodput of the various ACs with and without RD, 1 TCP Ack per
1 TCP Data segment, BER=0.}
\label{fig:ackonedata}
\end{figure}

In Figure~\ref{fig:ackonedata} (A)(B)(C)(D) 
we show the Goodput results for the BK, BE, VI and
VO ACs respectively, as a function of $K_D$.
These results are derived from Eq.~\ref{equ:goodput}
and were validated by simulation.
The simulation is carried out by a software that we wrote
and it is verified by analysis using a Markov chain model (in the
Appendix ).

In every graph there are curves for
$n=1,2,3,4,25$. There is also a curve for the Goodput
in the operation mode where RD is not used, i.e. the AP
and the station use the 'regular' IEEE 802.11ac MAC and
compete for the channel in every transmission attempt.
We denote this scenario by {\it No-RD}. This curve was
obtained by simulation
and the station always
tries to transmit as many MPDUs as it can (up to 64), that are
in its Transmission Queue at the time
it acquires the right to transmit.
The results for the {\it No-RD} scenario are also
validated by an analysis based on a Markov chain. In the
Appendix we present the Markov chain and its design.

We see in all the graphs that as the number of
transmissions increases and/or as $K_D$ increases,
so does the Goodput. We also include a curve showing
the maximum possible Goodput using RD. This curve is obtained
as follows: For every $K_D$ we first find the maximum number
of possible transmissions, $n_{max}$, such that
$n_{max} = \floor{\frac{178 \cdot 64}{7 \cdot K_D}}$.
Recall that $64 \cdot 178$ is the maximum number
of TCP Acks that the receiver can transmit in a TXOP.
Then, we compute the received Goodput for $n_{max}$
using Eq.~\ref{equ:goodput}. For example, for $K_D=64$ holds that $n_{max}=25$
and for $K_D=1$ holds $n_{max} = \floor{\frac{167 \cdot 64}{7 \cdot 1}}=1627$.

Notice that in the VI and VO ACs and for $K_D$s larger than 15, the
difference in performance between {\it No-RD} and using
RD is the largest among all the ACs. This happens because in these
ACs $CW_{min}$ and $CW_{max}$ are the smallest among the ACs
and so the probability for collisions is the largest.
In large $K_D$s collisions waste relatively long intervals of time
and so the decrease in the Goodput is significant. As $CW_{min}$ and
$CW_{max}$ decrease, the difference between using RD and {\it No-RD}
increases. Notice that in VO $CW_{min}=4$ and so the
collision probability is $25\%$. In VI $CW_{min}=8$ and
the collision probability is $12.5\%$. For the BK and BE
$CW_{min}=16$ and the collision probability is only $6.25\%$.

For smaller $K_D$s, i.e. $1 \le K_D \le 15$, notice that
{\it No-RD} sometimes outperforms $RD(1)$. In BK and BE
the collision probability is small and the AP and the station
transmit almost alternately. Therefore, {\it No-RD} and $RD(1)$
have almost the same performance, except that in $RD(1)$ there
is an extra overhead of {\it CF-End} and $SIFS$ at the
end of every TXOP. 

As the value of AIFS is larger, this overhead is less
significant. In BE the AIFS is $43 \mu s$ compared to
$79 \mu s$ in BK and therefore
the {\it CF-End}$+SIFS=26+16=48 \mu s$ is more significant
in BE and {\it No-RD} slightly outperforms $RD(1)$, while
in BK they perform equally. 

In VI and VO the AIFS is smaller
than in BK and BE and so the {\it CF-End}$+SIFS$ overhead is
more significant. Moreover, the AP in these ACs has
a higher probability of accessing the channel
than the station because its AIFS is shorter by
one {\it Slot-Time}. This enables the AP
in {\it No-RD} to transmit several times in a row
before the station replies. This also
enables a better Goodput in {\it No-RD} than in $RD(1)$
where the AP and the station transmit alternately. 
On the other hand, the collision probability 
is larger in VI and VO. However, the AP transmits
many times without competition in {\it No-RD}
when the TCP receiver has no TCP Acks to transmit.
The overall outcome is a slightly larger Goodput
in {\it No-RD}, compared to $RD(1)$, than in BK and BE.

\begin{figure}
\vskip 16cm
\includegraphics{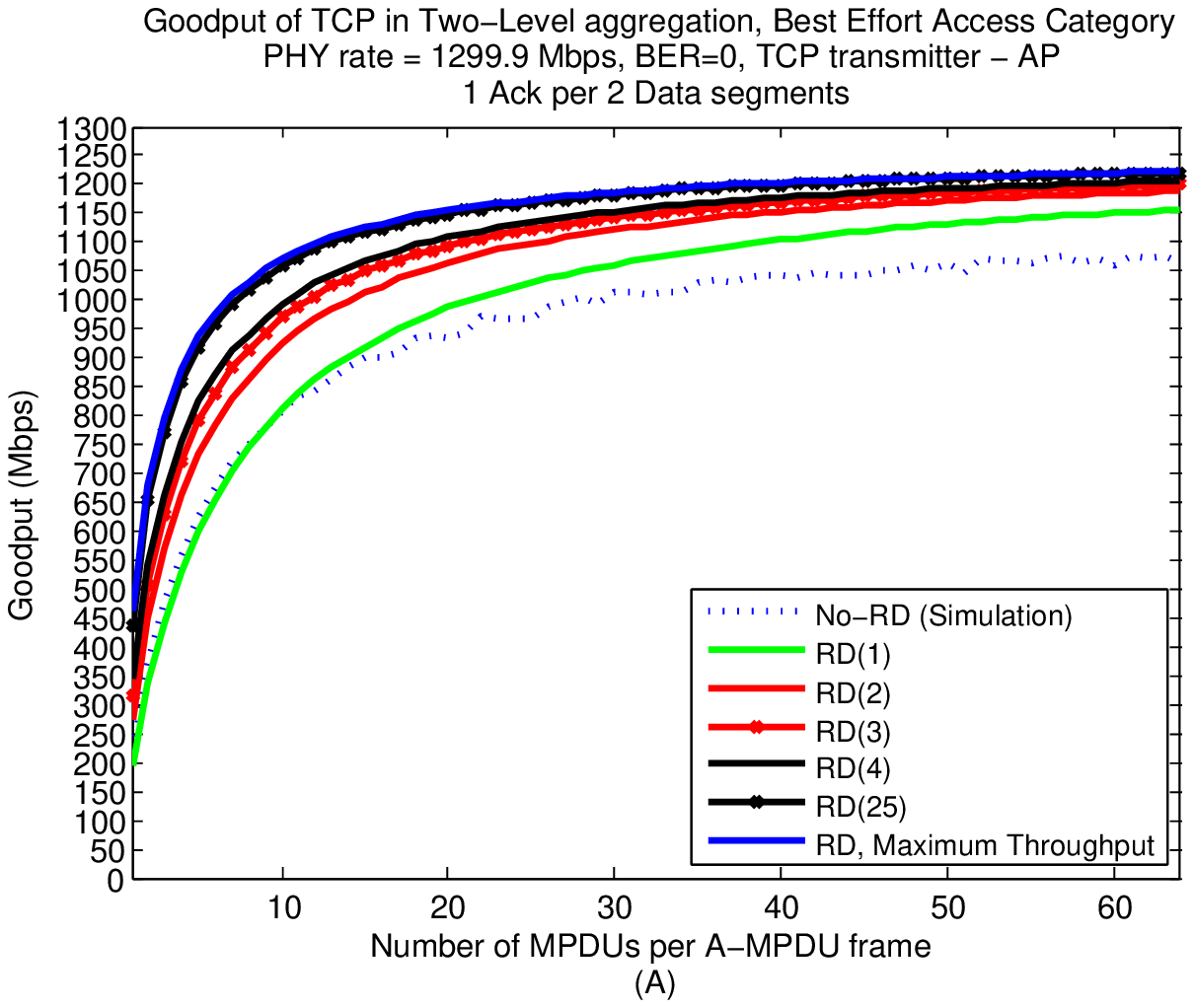}
\includegraphics{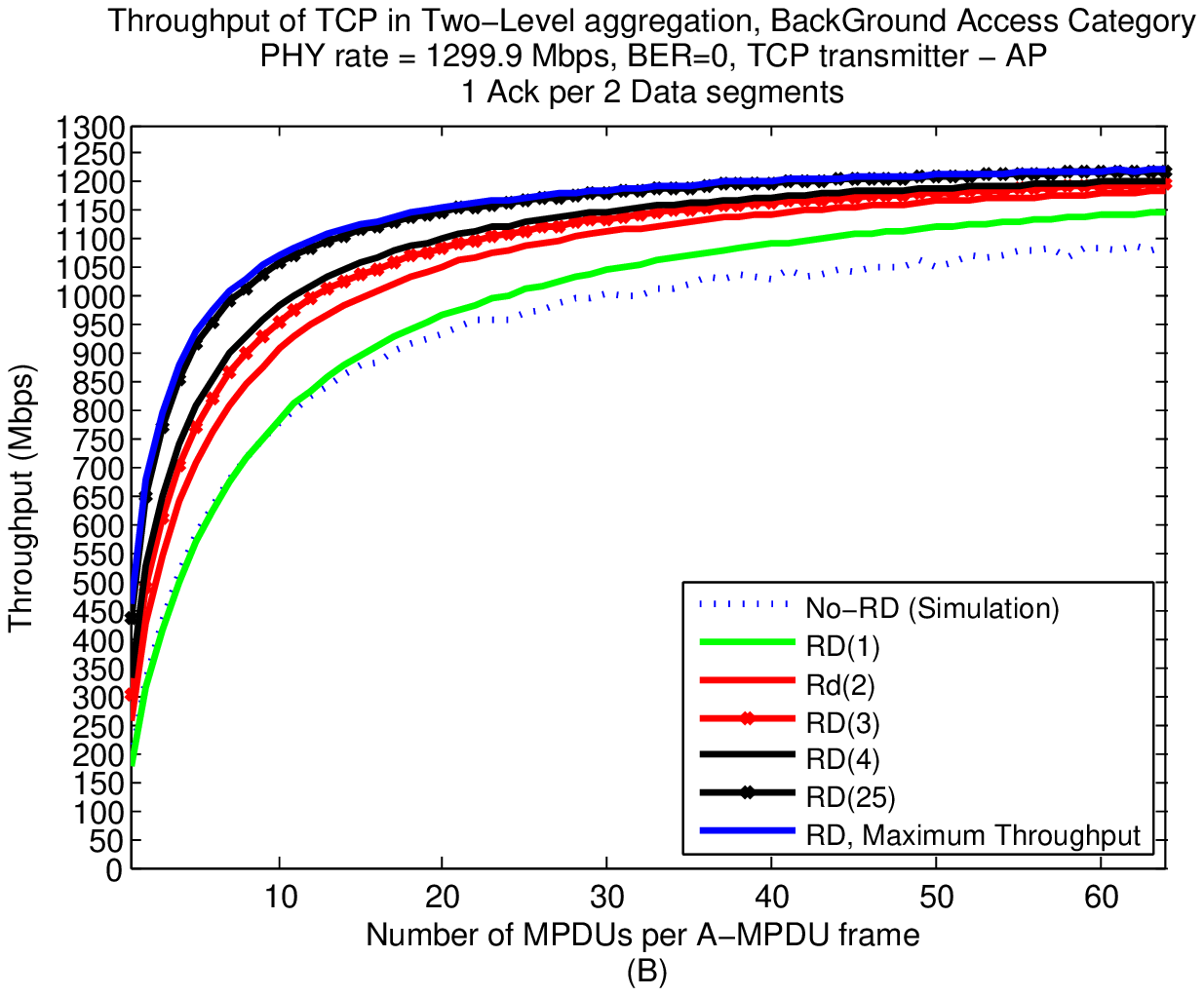}
\includegraphics{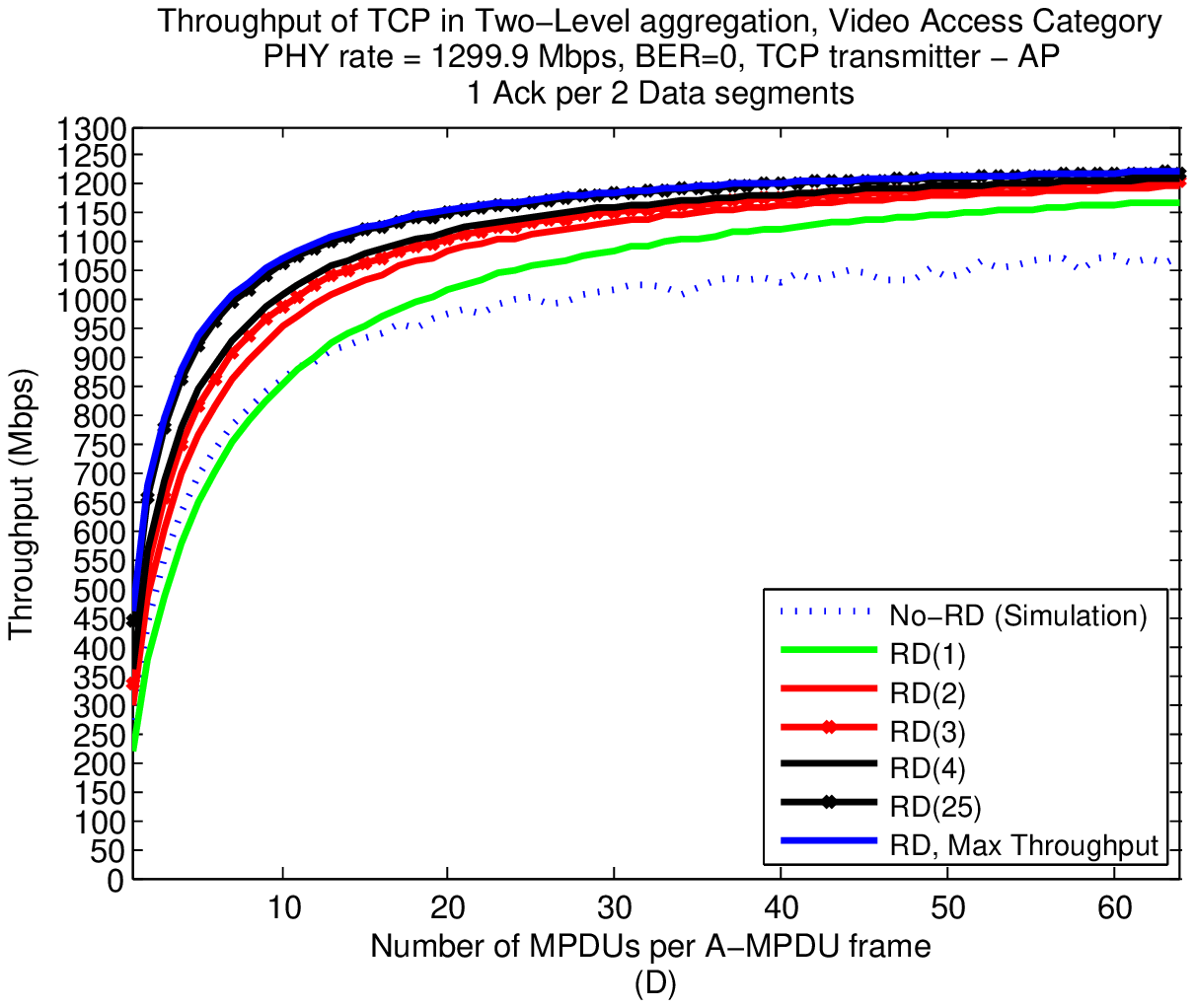}
\includegraphics{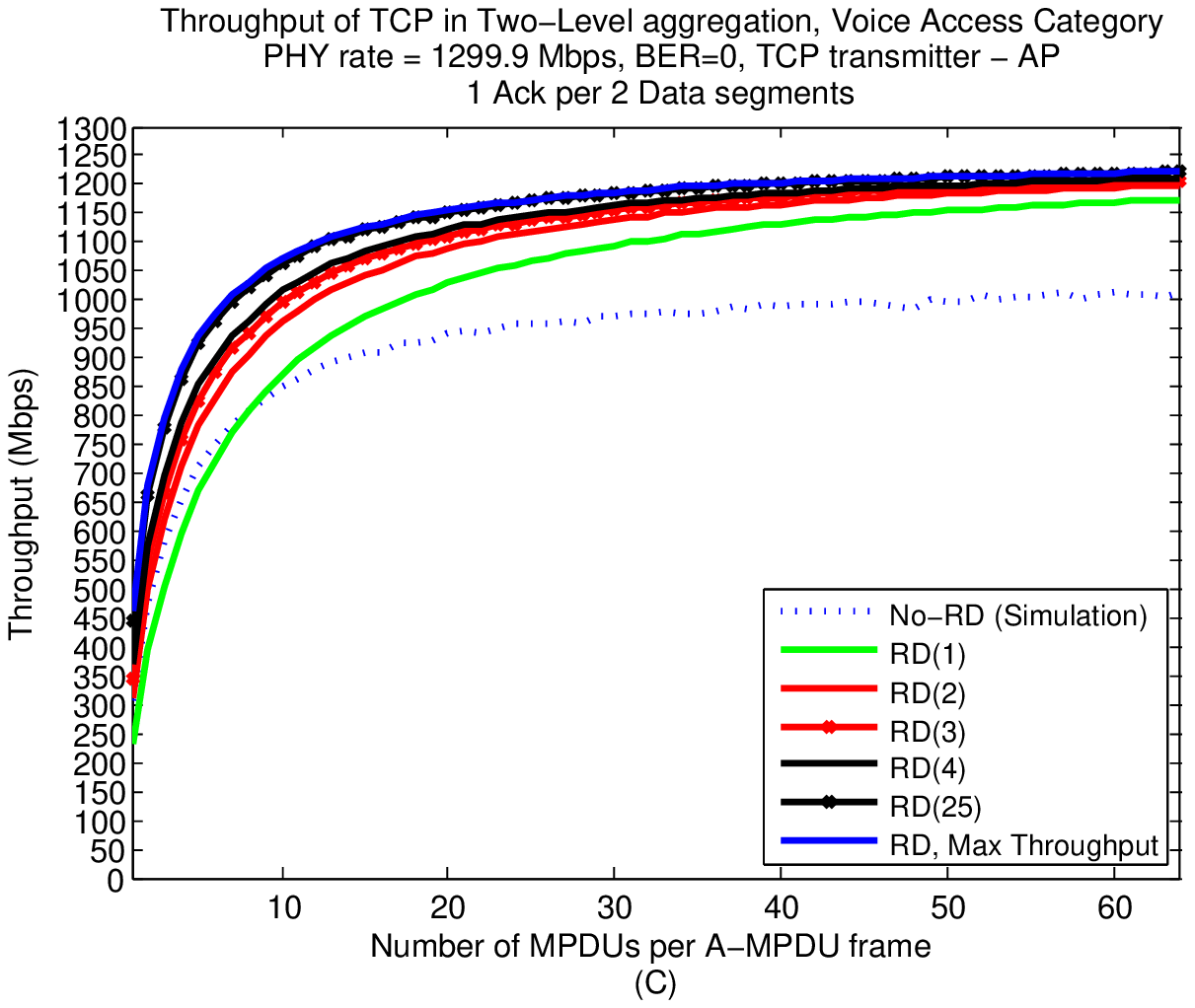}
\caption{The Goodput of the various ACs with and without RD, 1 TCP Ack per
2 TCP Data segment (TCP Delayed Acks), BER=0.}
\label{fig:acktwodata}
\end{figure}

In Figure~\ref{fig:acktwodata}  
we show the same results as in Figure~\ref{fig:ackonedata}
but now every TCP Ack acknowledges two TCP Data segments,
a feature known as {\it TCP Delayed Acks}. 
For clarity, for the {\it No-RD} scheme
we only show the simulated results. The analytical
results are similar, as can be seen in Figure~\ref{fig:ackonedata}.
Normally, the TCP receiver does not send an Ack
the instant it receives data. Instead, it delays the
Ack, hoping to have data going in the same direction
as the Ack, so the Ack can be sent along with the data.
This delay is usually in the order of $200 \mu s$.
However, if meanwhile another data segment arrives, the
TCP receiver immediately generates an Ack to send.

Using TCP Delayed Acks
enables the TCP transmitter to transmit more TCP
Data segments in one TXOP : the limiting condition is
now $n \cdot K_D \cdot 7 \le 2 \cdot 64 \cdot 178$.
Comparing between the curves of the Maximum Goodputs
in the cases of with and without TCP Delayed Acks
reveals 
an improvement of only about $2\%$ in the Goodput for
large $K_D$s in the case of
TCP Delayed Acks.
The reason for the small
improvement can be understood from Eq.~\ref{equ:goodput}:
the main impact of TCP Delayed Acks is in enabling more
transmissions of the TCP transmitter during a TXOP.
However, increasing the number of transmissions $n$ does
not increase the Goodput significantly for 'large' $n$'s.

\begin{figure}
\vskip 8cm
\includegraphics{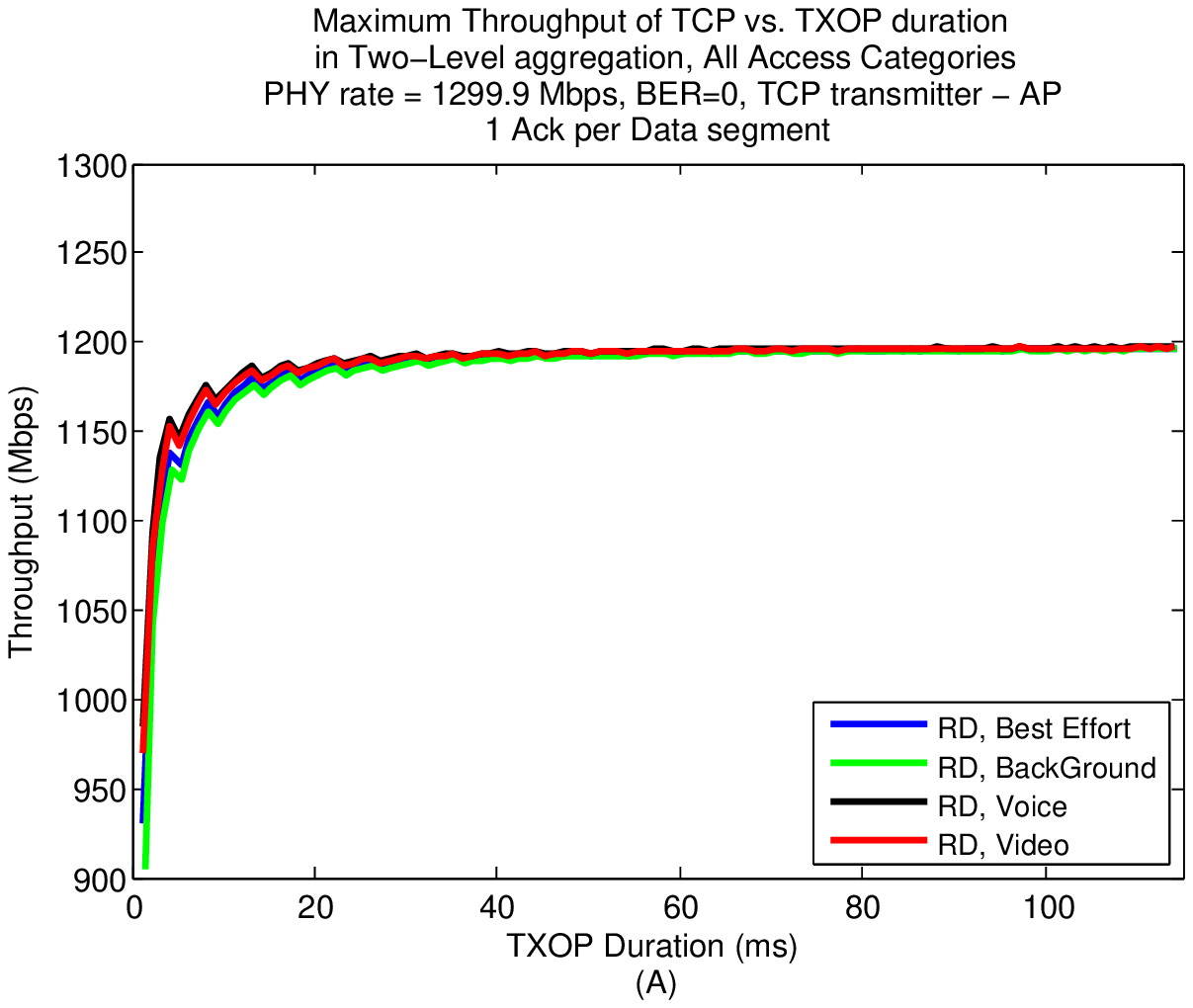}
\includegraphics{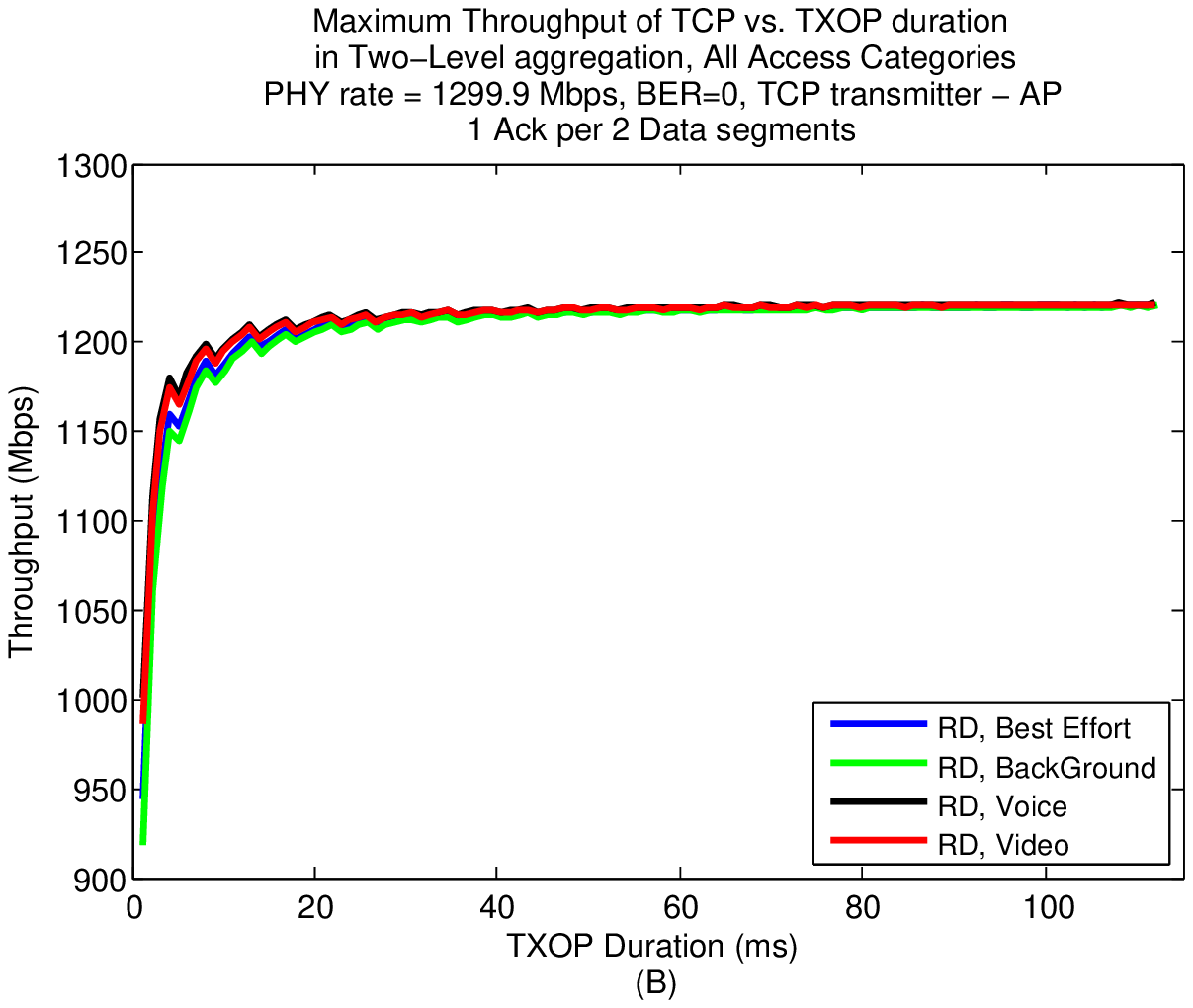}
\caption{The maximum Goodput vs. the duration of the TXOPs in the various ACs,
with and without TCP Delayed Acks, BER=0. }
\label{fig:txop}
\end{figure}

Finally, in Figure~\ref{fig:txop} we show the Goodput of the
various ACs as a function of the TXOP duration.
In Figures~\ref{fig:txop} (A) and (B) we assume 
that TCP Delayed Acks are not in use and in use respectively.
The curves were computed as follows: For every $S$ TCP
Data segments, $1 \le S \le 64 \cdot 178 $ ( $2 \cdot 64 \cdot 178$ for
TCP Delayed Acks ) we explored the most
efficient way ( Maximum Goodput ) to transmit these 
segments, and the achieved Goodput.
The most efficient way is to transmit as many MPDUs
as possible in a single transmission, up to
64, while every MPDU contains 7 MSDUs except
the last one, when $S$ is not divided by 7.
We then checked the cycle length, and how the curves
relate between cycles' lengths and Goodputs.
For example, if
we find that transmitting $S$ TCP Data segments
with the largest Goodput $G$ takes $C ms$, it is easy
to verify that $G$ is the largest Goodput possible
in $C ms$.

We see that all the ACs achieve the same Goodput
for 'long' TXOP. This happens because the
cycles in the various ACs differ only in the
AIFS and BackOff time intervals which become
negligible in long cycles.
In shorter cycles the VI and VO ACs achieve
the same best performance because their
AIFS are the shortest, $25 \mu s$ for the AP.
BE outperforms BK because its AIFS is $43 \mu s$ (AP)
compared to $79 \mu s$ in BK. See Table 1.

There are two
important outcomes from Figure~\ref{fig:txop}.
First, using a TXOP of 20-30$ms$ is sufficient to achieve
almost the largest Goodput possible. This is important
since it enables short time-outs in the TCP protocol
and so the TCP transmitter can receive TCP Acks
sooner, while still using the wireless channel efficiently. 
Second, in a scenario where there are several TCP
connections between the AP and several stations, it is
sufficient for the AP to use TXOPs of 20-30$ms$ in order
to use the channel efficiently. This has an impact
on the fairness among the stations and is the basis for
further research.

\section{Error-Prone channel Results}

\begin{figure}
\vskip 16cm
\includegraphics{gbe1.ps}
\includegraphics{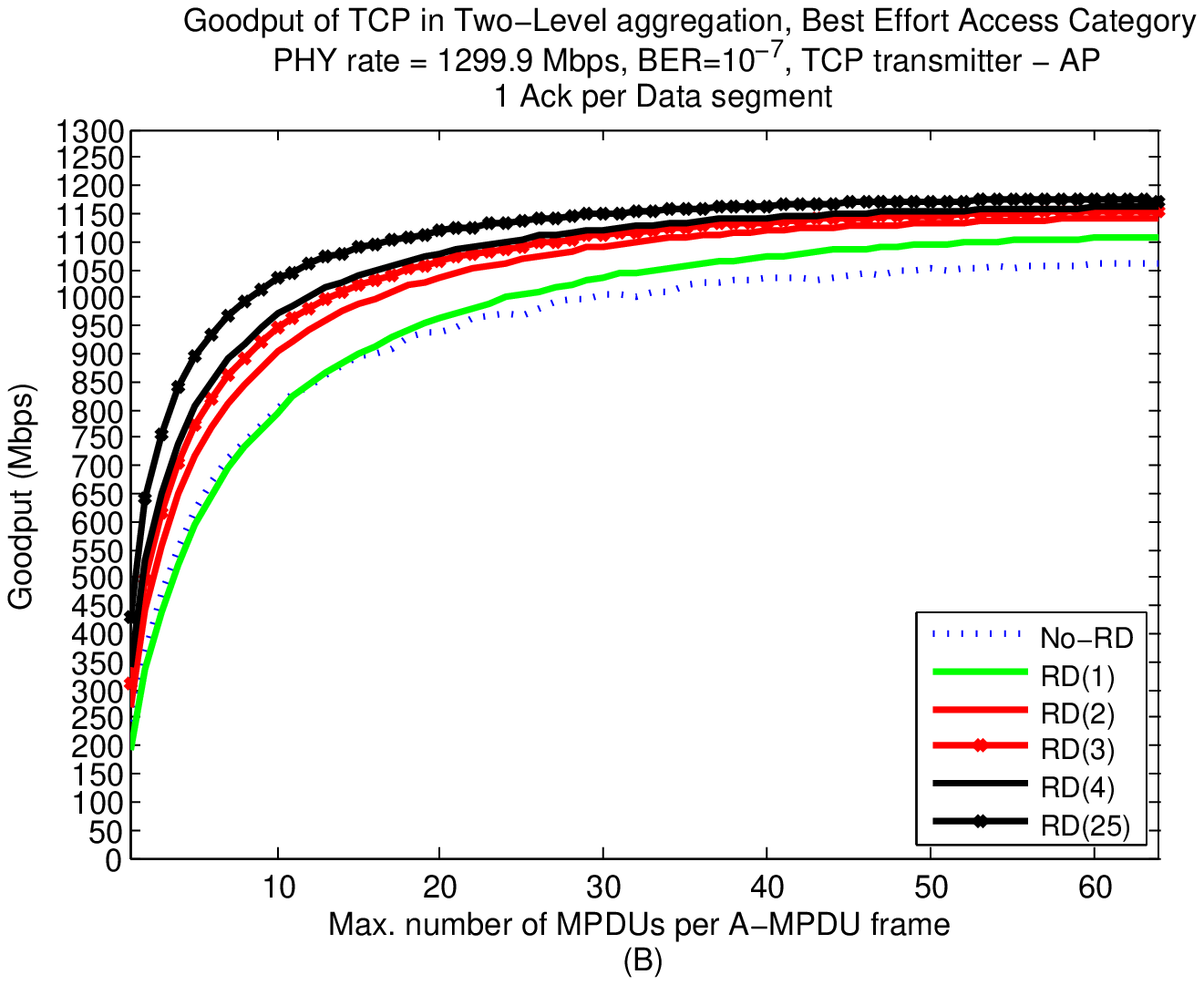}
\includegraphics{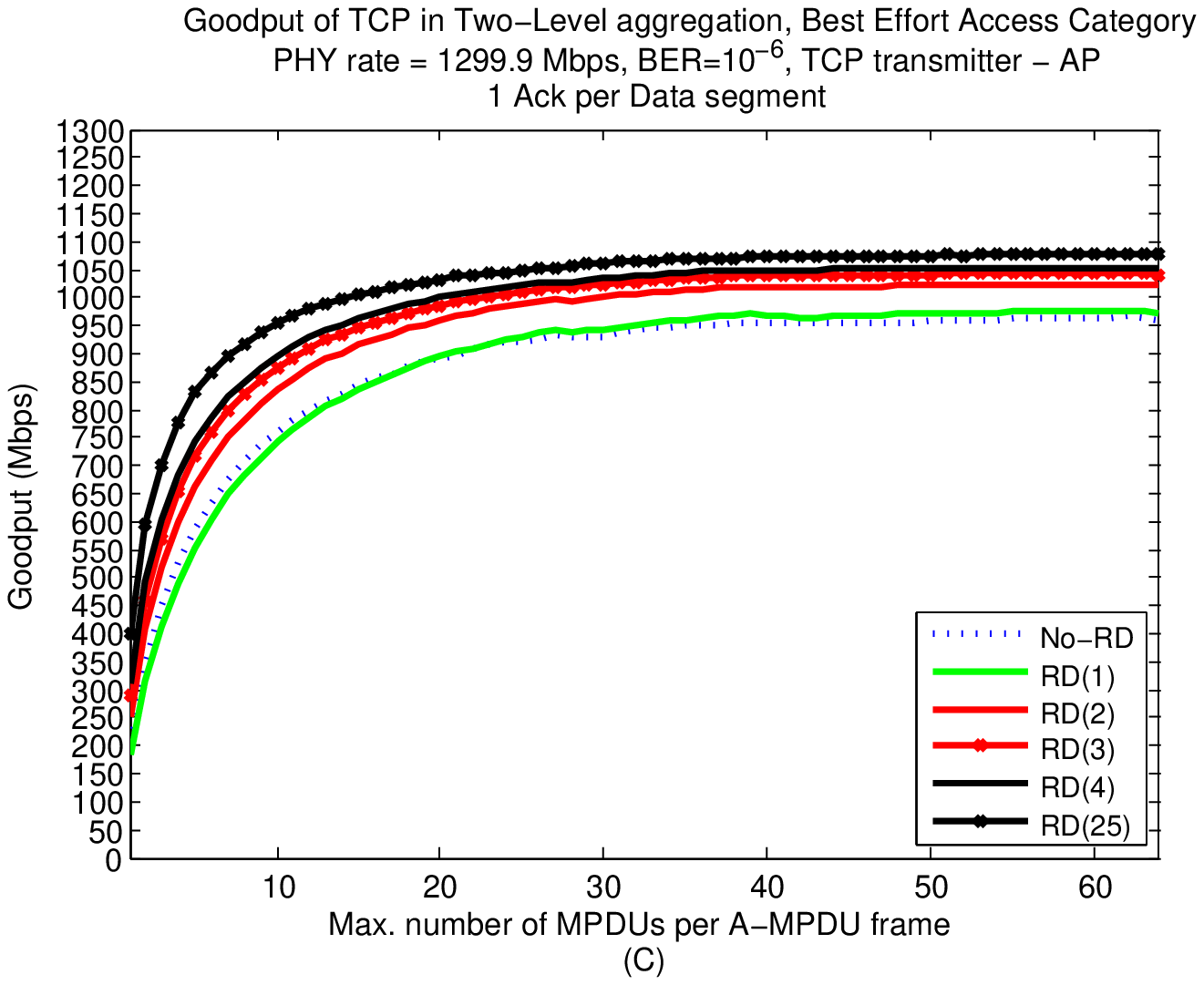}
\includegraphics{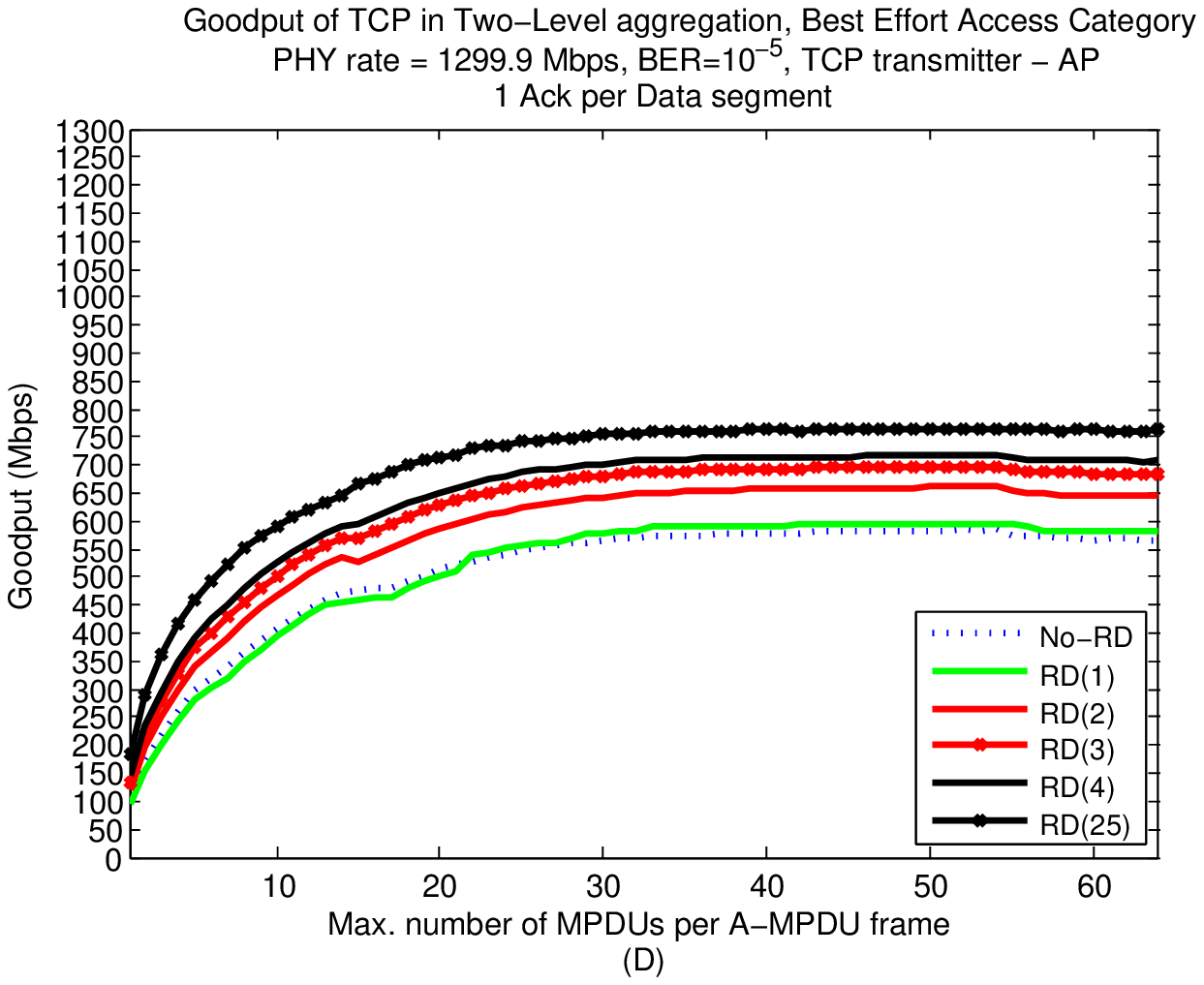}
\caption{The Goodput of the Best Effort AC with and without RD for various BERs, 1 TCP Ack per 1 TCP Data segment.}
\label{fig:be1allber}
\end{figure}

In this Section we assume the BERs of $10^{-5}, 10^{-6}$ and $10^{-7}$.
We concentrate only on the BE AC. The results
for the other ACs are similar, with the same differences compared
to BE as described in Section 4.
In Figure~\ref{fig:be1allber} (A),(B),(C),(D)
we show the Goodput vs. the
maximum number $K_D$
of MPDUs per transmission of the TCP transmitter in the
BE AC for BER$=0, 10^{-7}, 10^{-6}, 10^{-5}$ respectively.
First notice that the title of the X-axis in Figure~\ref{fig:be1allber}(A)
is different than those of parts (B),(C) and (D). This is
because the positive BER can cause the MAC TW to
limit the number of transmitted MPDUs in a single
transmission to be smaller than $K_D$, and
so $K_D$ is only the maximum allowed MPDUs
in a single transmission.

In general using RD results in a larger Goodput.
Notice however that as the BER increases, the advantage
of $RD(1)$ over {\it No-RD} decreases. As the BER increases,
the number of MPDUs that the TCP transmitter is able
to transmit in every transmission decreases. The
MAC TW is not always able to slide so that
it will contain $K_D$ MPDUs, i.e. the maximum allowed 
number of MPDUs. This results
in two outcomes: First, as the BER increases a smaller
number of MPDUs are
transmitted in {\it No-RD} and $RD(1)$, and a smaller number of
MPDUs arrive successfully at the TCP receiver in both 
schemes. However, shorter A-MPDU frames have an
advantage in {\it No-RD} because
the penalty
of collisions is smaller.
These two outcomes cause
$RD(1)$ and {\it No-RD} to coincide as the BER increases.

In $RD(2)$ there is a second transmission in every TXOP
which increases the probability that MPDUs arrive
successfully. Thus the MAC TW slides
faster, enabling more successful transmissions of MPDUs.
This causes a significant improvement in the Goodput
of $RD(2)$ compared to {\it No-RD}.

Notice that Figure~\ref{fig:be1allber}(A) is for BER$=$0 
and it is the same as Figure~\ref{fig:ackonedata}(A). 
In Figure~\ref{fig:be1allber}(A) we can provide a curve showing
the maximum possible Goodput. 
However, for BER$>$0, in order to find such a curve
one needs to know, given $K_D$,
the actual average number of transmitted MPDUs in
every transmission of the TCP transmitter. This
number might be smaller than $K_D$, especially for large
$K_D$s, because it is possible that the 
MAC TW
does not contain $K_d$ MPDUs.  Such a computation is
difficult~\cite{GK,SA1} and it is out of the scope of this paper.
This is also the reason why we cannot provide
analytical results for the {\it No-RD} scheme as for
the case BER$=0$.
Notice again that for small $K_D$s {\it No-RD} slightly outperforms
$RD(1)$ for the same reasons given for this phenomena in Section 4.

\begin{figure}
\vskip 16cm
\includegraphics{gbe2.ps}
\includegraphics{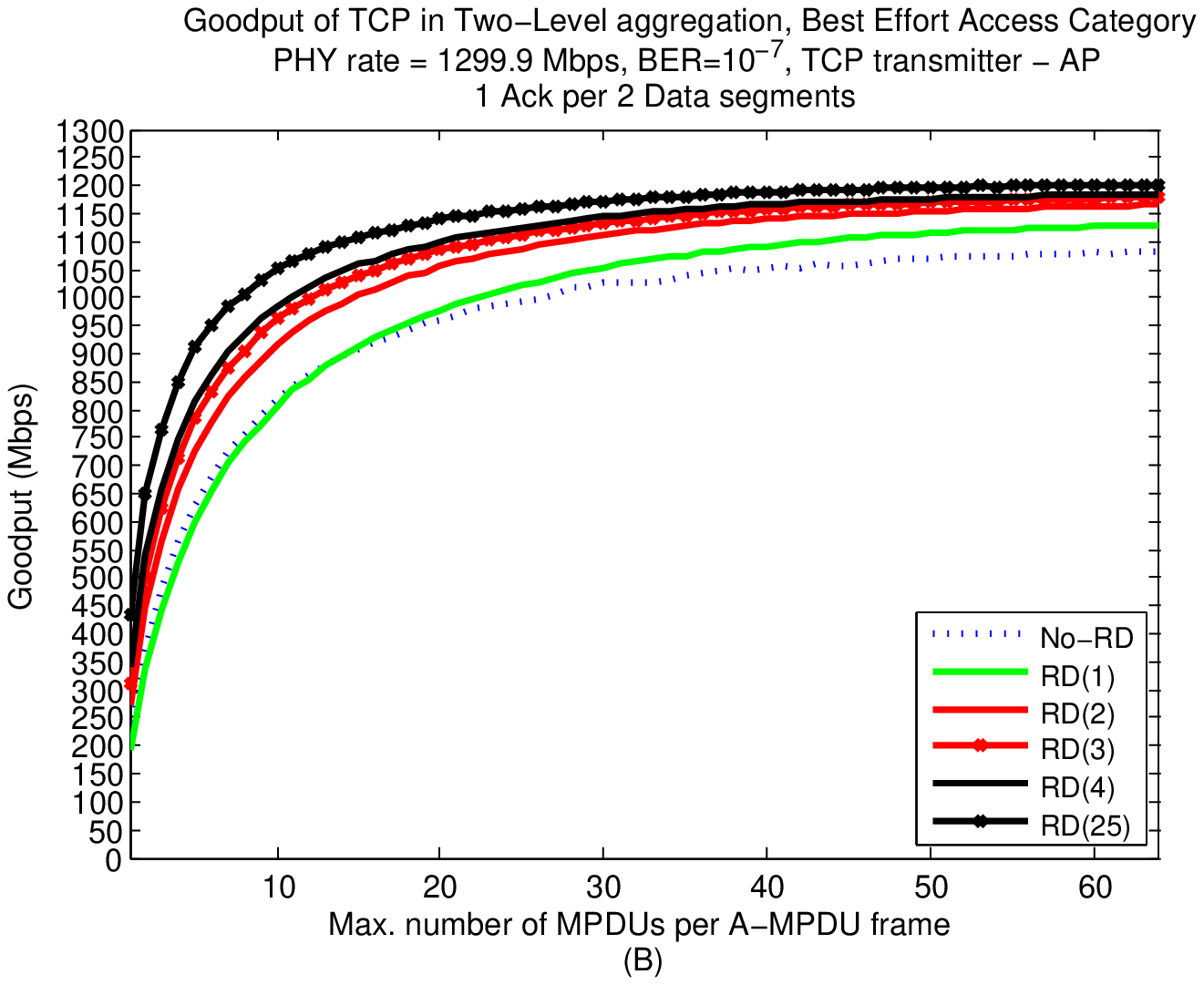}
\includegraphics{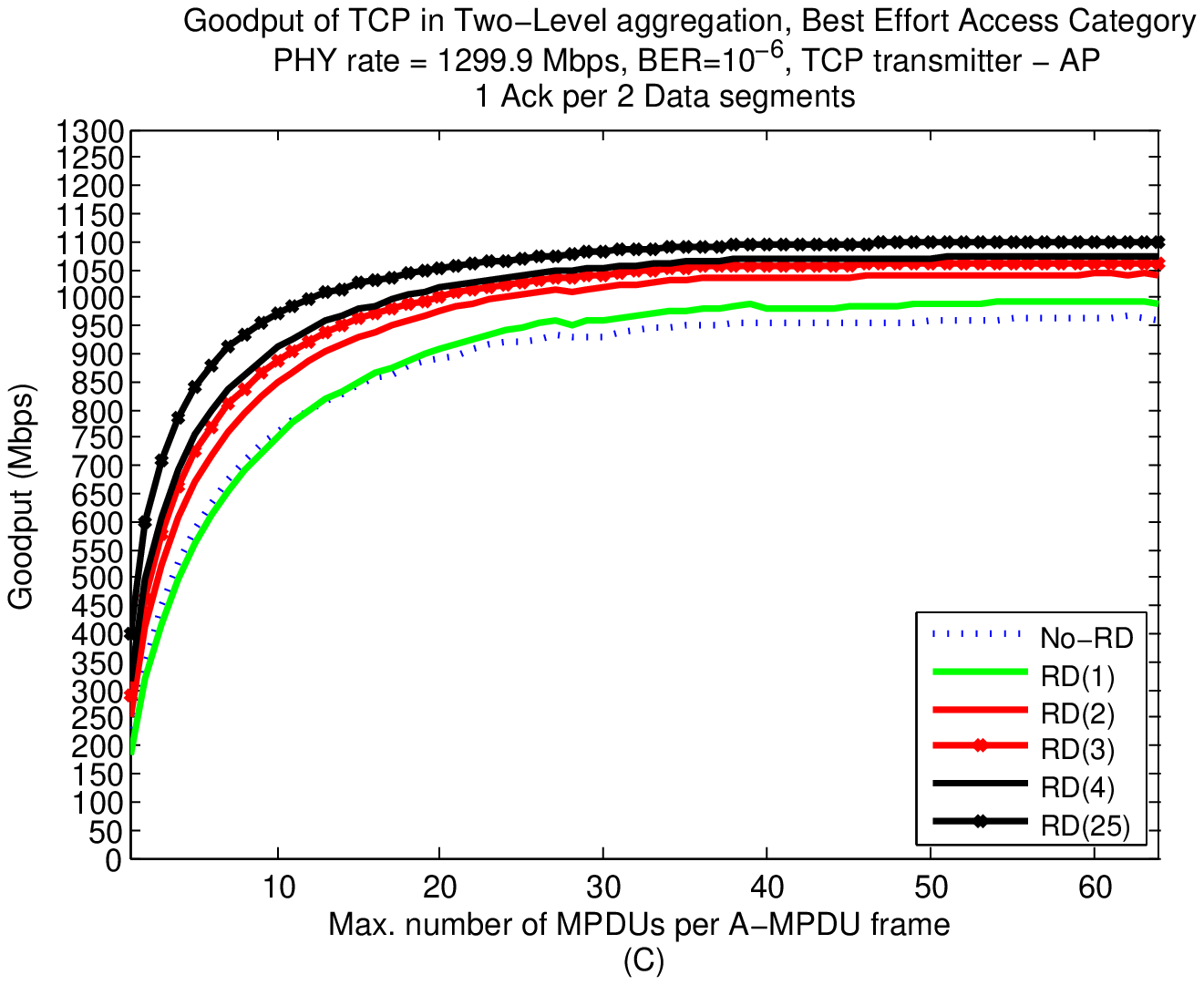}
\includegraphics{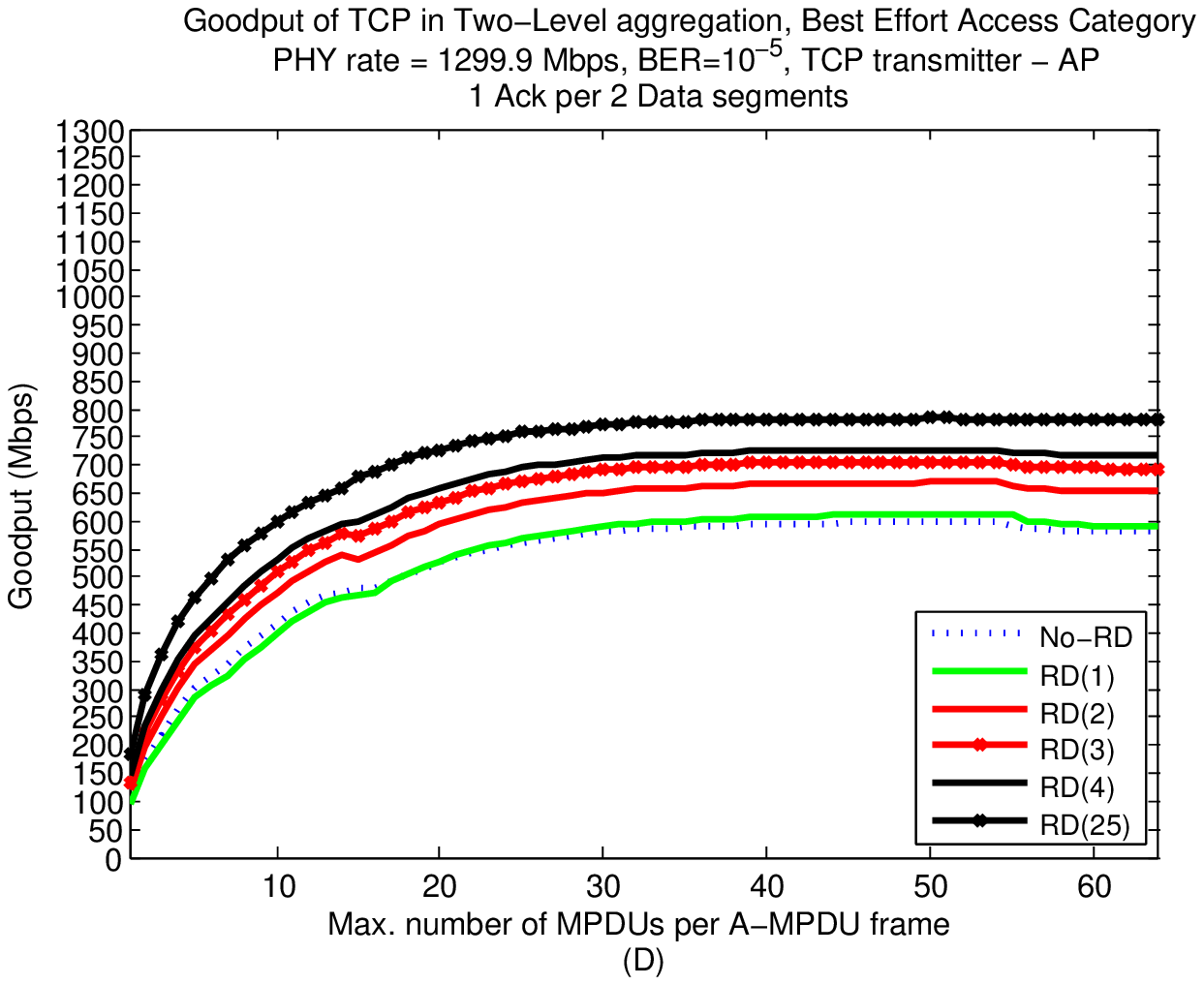}
\caption{The Goodput of the Best Effort AC with and without RD for various BERs, 1 TCP Ack per 2 TCP Data segments (TCP Delayed Acks).}
\label{fig:be2allber}
\end{figure}

In Figure~\ref{fig:be2allber} we show the same results as in 
Figure~\ref{fig:be1allber}, but now there is a use in the
TCP Delayed Acks. Using TCP Delayed Acks does not
improve the performance of the 
{\it No-RD} scheme because 
in the case of collisions, the time
wasted is the time of transmitting the TCP Data segments.
The shorter time of transmitting the TCP Acks has no influence
in this case. On the other hand, in the schemes that use RD the reduced
time of transmitting TCP Acks has an influence because 
the TXOP length is shorter. Therefore, one can see
that the difference between the performance of the RD schemes
to that of {\it No-RD} is larger than in the case of not using
TCP Delayed Acks.

\begin{figure}
\vskip 16cm
\includegraphics{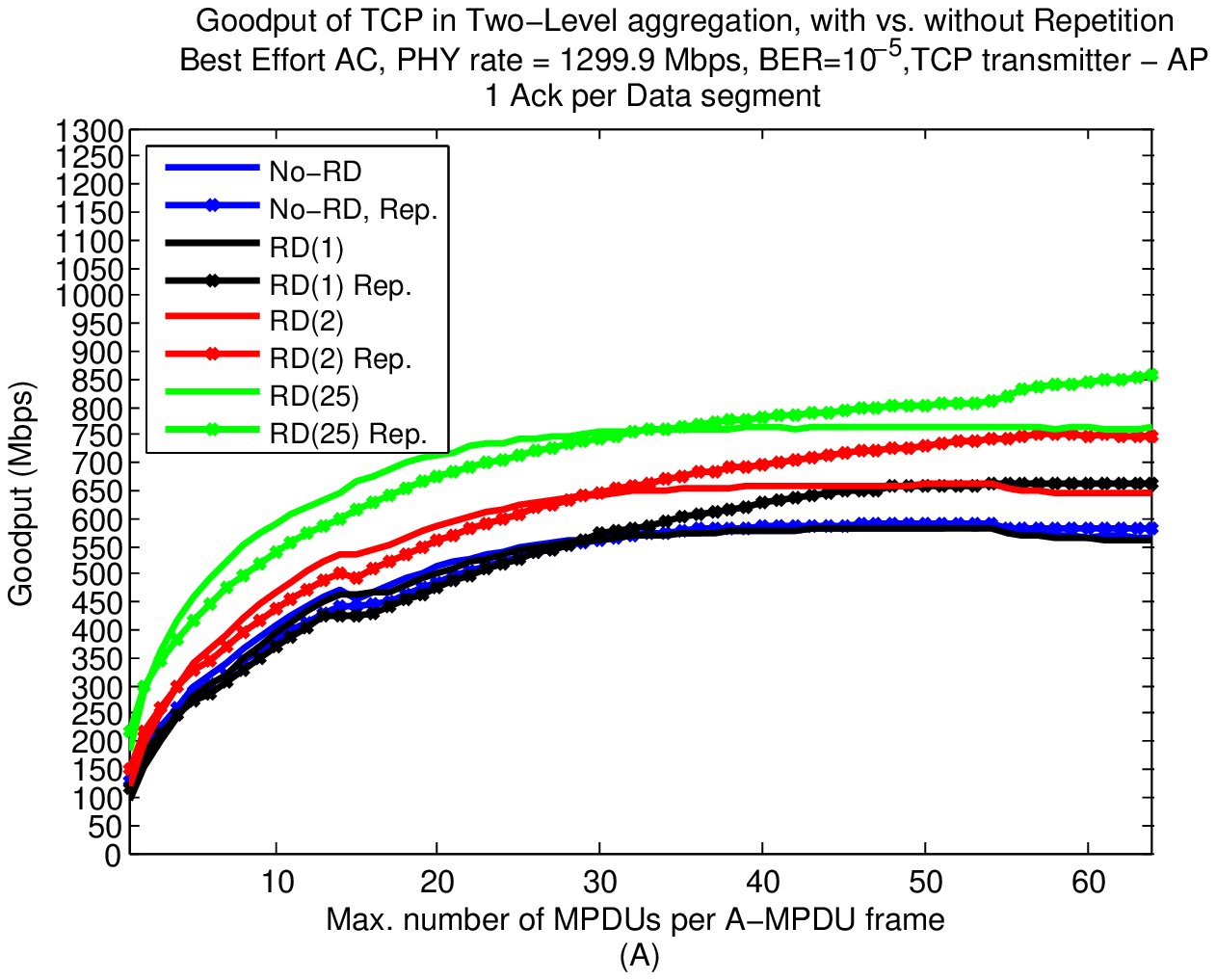}
\includegraphics{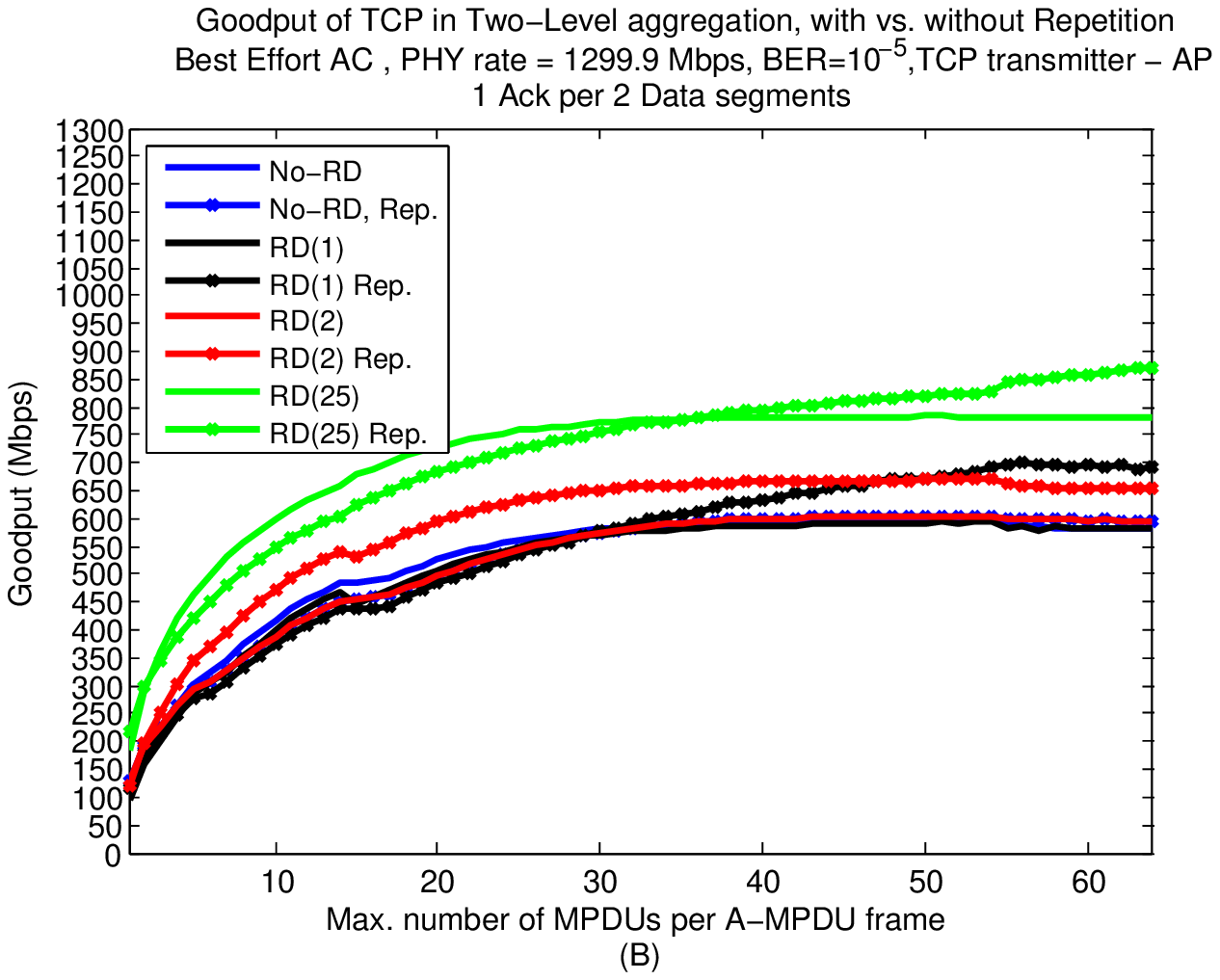}
\includegraphics{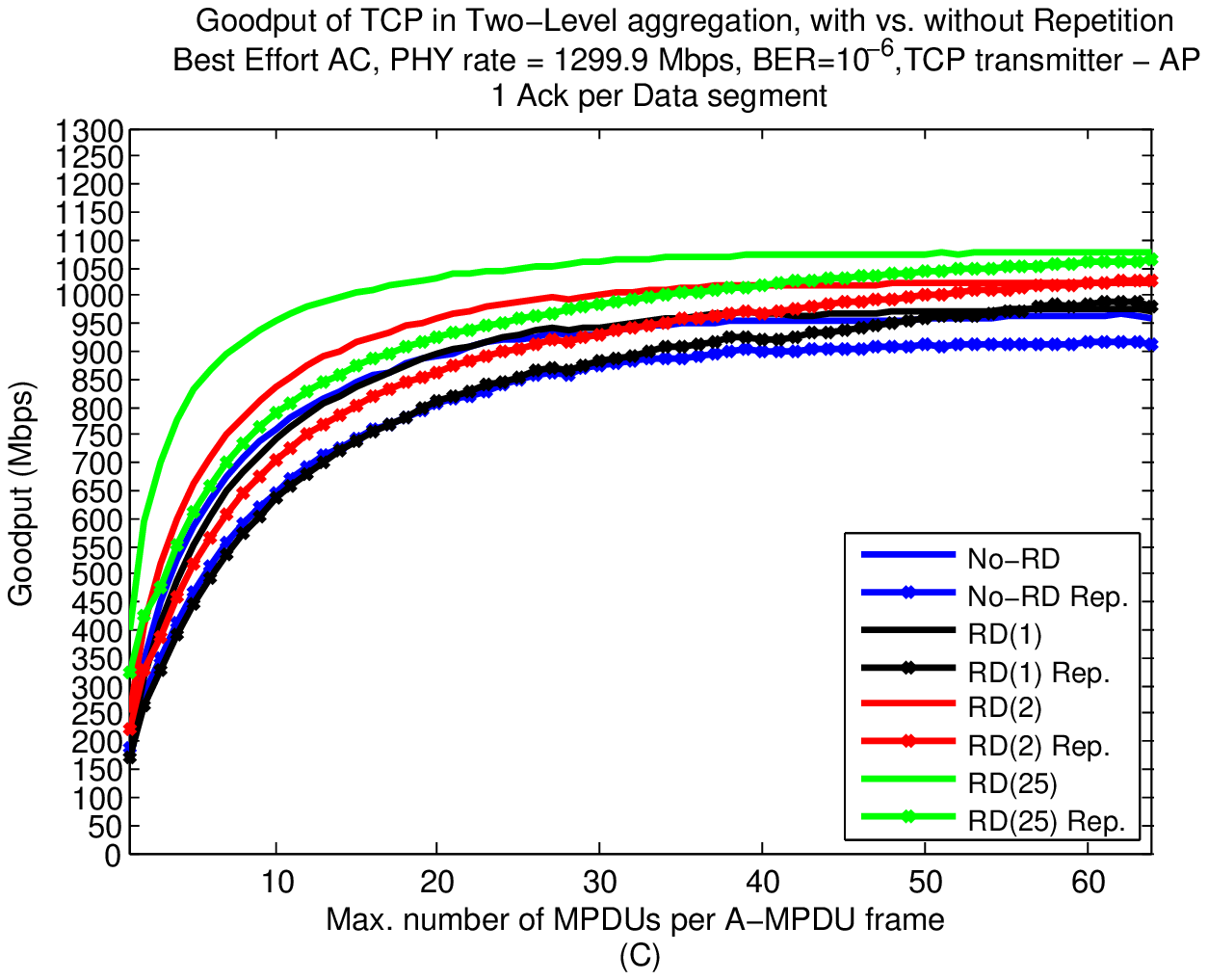}
\includegraphics{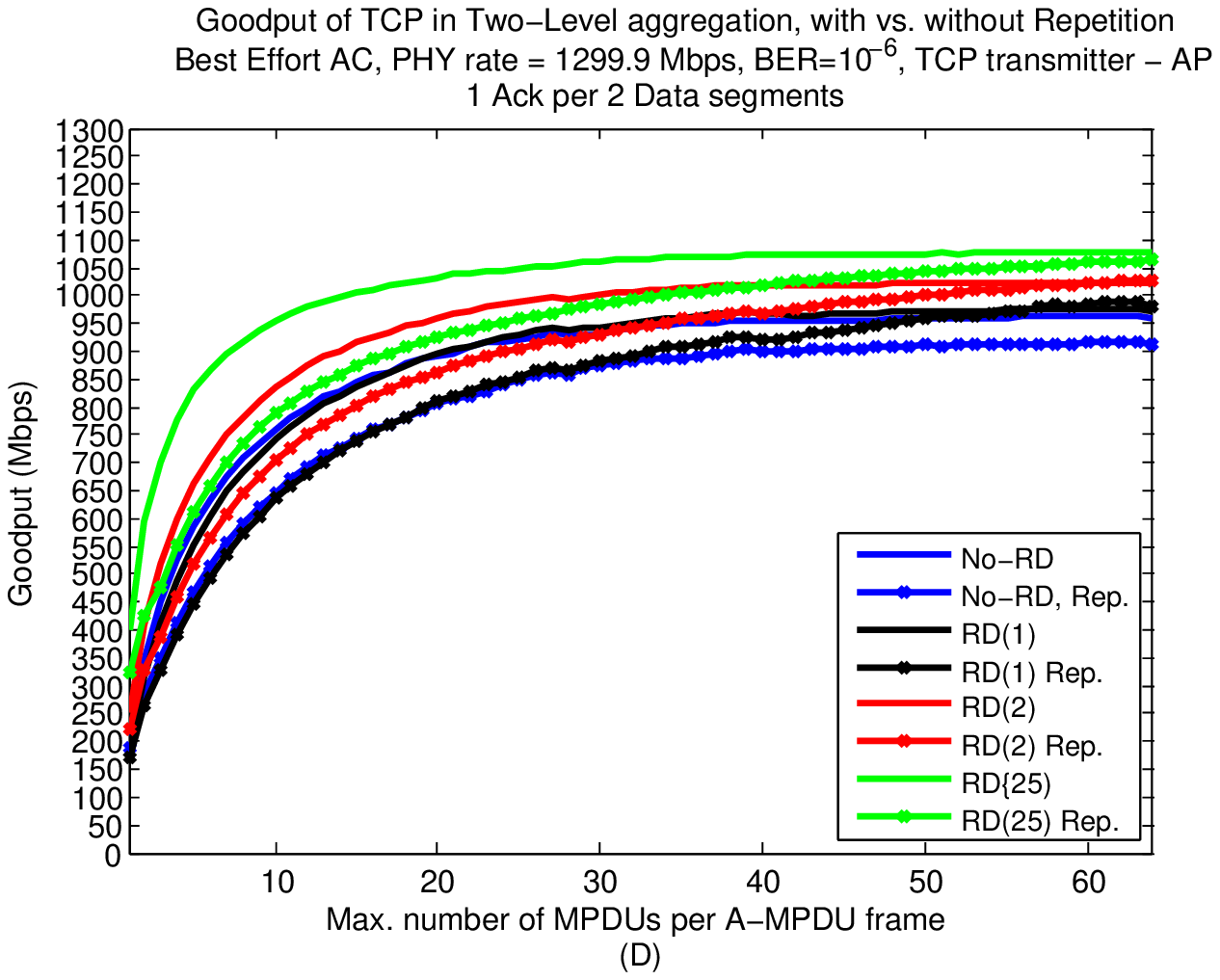}
\caption{Comparing transmissions with and without $Rep.$ in the Best Effort AC.}
\label{fig:mix56}
\end{figure}

In Figure~\ref{fig:mix56} we show the use in the scheme
of~\cite{SA2} where each of the first 3 MPDUs
in every A-MPDU frame of the
TCP transmitter is transmitted twice, i.e. MPDU repetition. 
Only the first 3 MPDUs are
transmitted twice because it is the most
efficient scheme (max. Goodput), as was shown in~\cite{SA2}.
This scheme, which we denote $Repetition (Rep.)$, has two
effects. First, it increases the arrival success probabilities
of the first 3 MPDUs in the A-MPDU frames.
As an outcome, it enables the MAC TW
to slide faster and to contain
more MPDUs ready for transmission, compared to
the case of not using $Rep$. On the negative side the
transmission time of the A-MPDU frames increases by transmitting
the first 3 MPDUs twice. Therefore, for BER$=$0 it is clear
that the performance of $Rep$. is worse than not using it.
As the BER increases, the advantage of $Rep$. increases.
We also found that for BER$=10^{-7}$ it is inefficient
to use $Rep$. However, for BER$=$$10^{-5}, 10^{-6}$ $Rep$.
improves the achieved Goodput as we show
in Figure~\ref{fig:mix56}.

In order to demonstrate the improvement consider
Figure~\ref{fig:mix56}(A) for BER$=10^{-5}$ without
TCP Delayed Acks.
One can see that
all the schemes, namely {\it No-RD, RD(1), RD(2)} and $RD(25)$,
benefit from using $Rep$. in the case of large $K_D$s,
while for small $K_D$s it is not efficient.
Notice that in the case of small $K_{D}$s the
probability that the MAC TW
will contain $K_{D}$ MPDUs ready for transmission
is much larger than in the case of larger $K_{D}$s.
Therefore using $Rep$. in the former case
only increases the transmission time of the
TCP transmitter A-MPDU frames with no benefit.

\begin{figure}
\vskip 8cm
\includegraphics{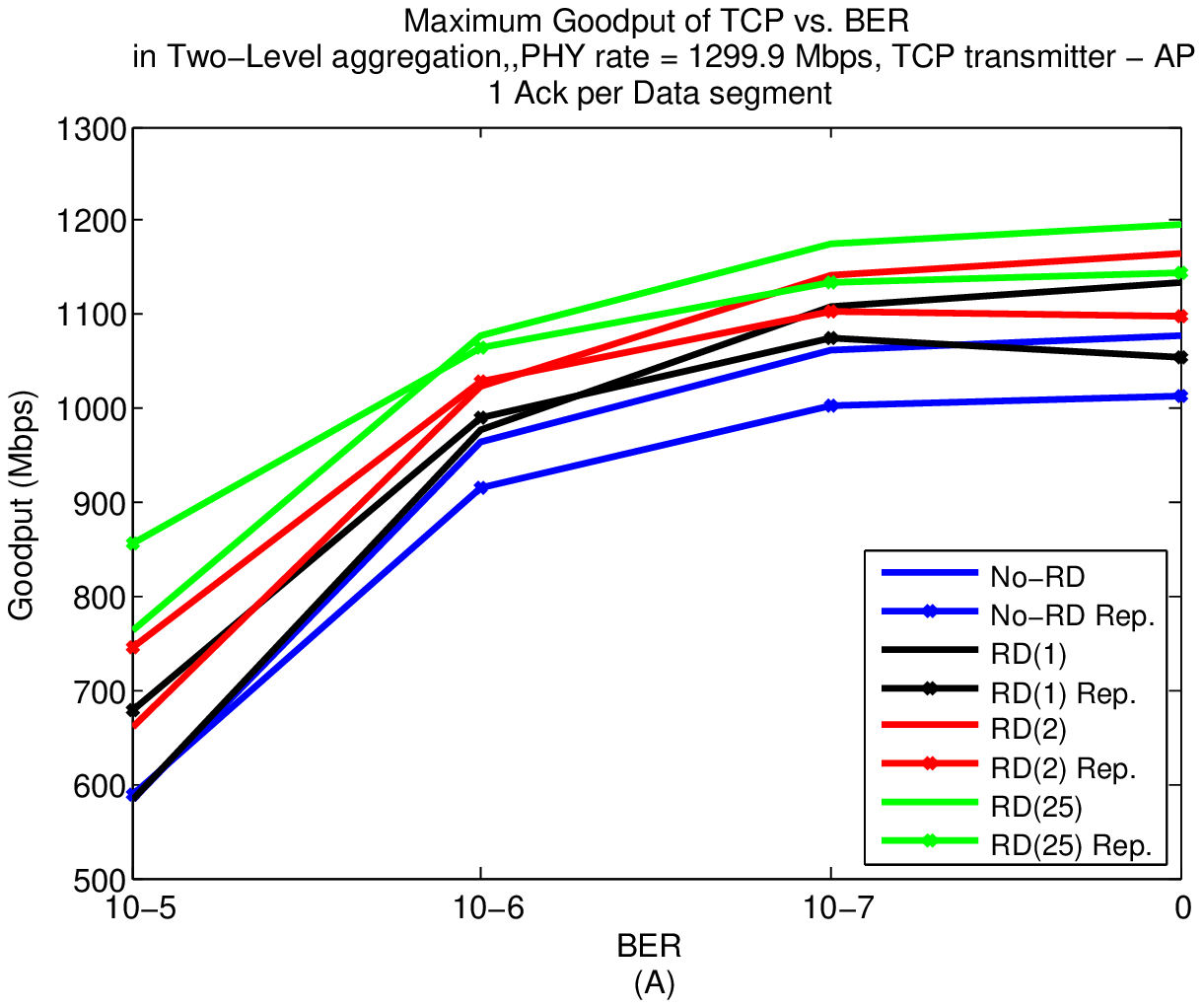}
\includegraphics{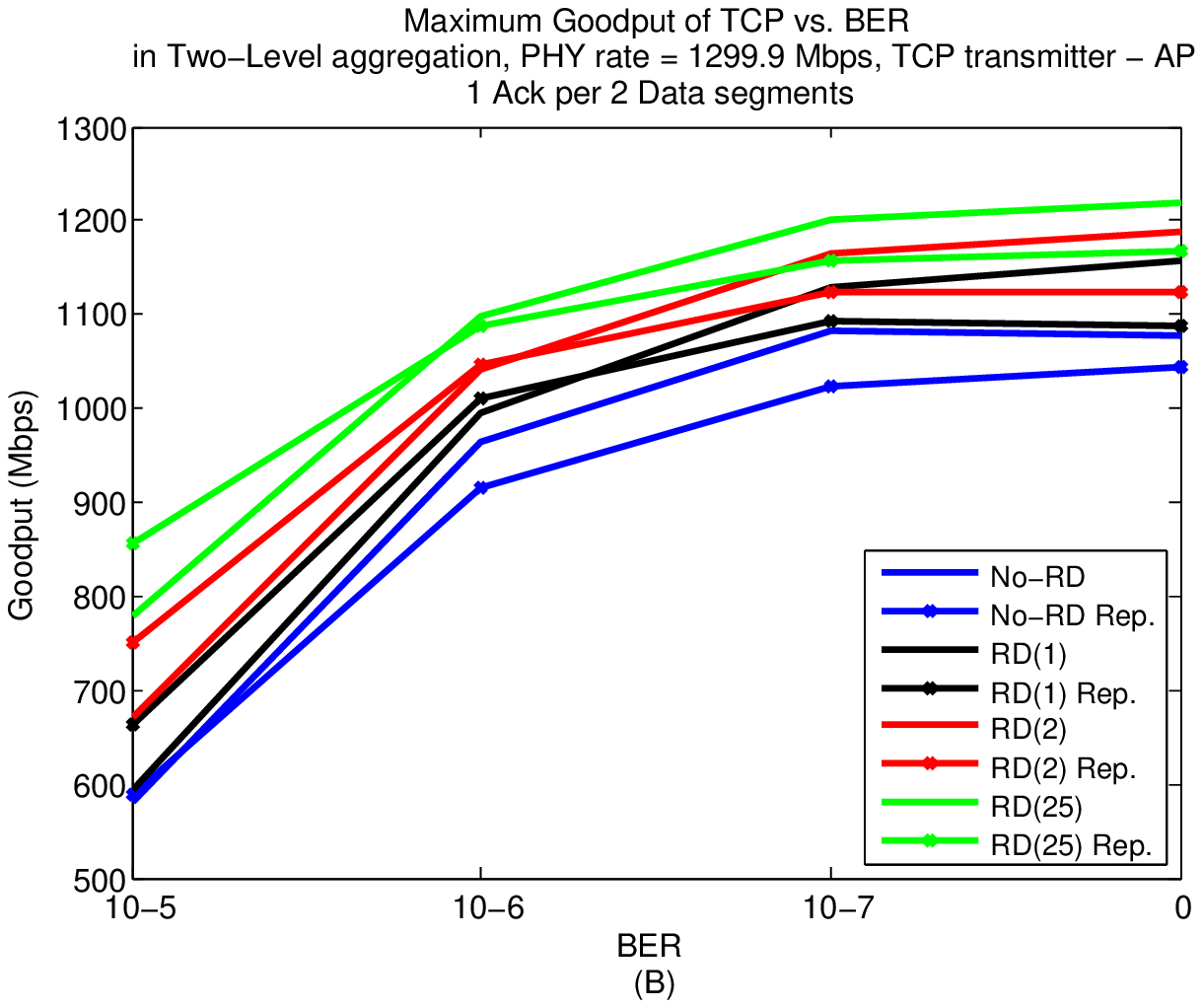}
\caption{The maximum Goodput in the various transmission schemes vs. BER,
for the Best Effort AC.}
\label{fig:maxthrbe}
\end{figure}

In Figure~\ref{fig:maxthrbe} we show the maximum received 
Goodputs vs. the BER for the {\it No-RD, RD(1), RD(2)} and $RD(25)$ schemes.
In Figures~\ref{fig:maxthrbe}(A) and (B) we 
consider the cases without and with TCP Delayed Acks respectively.
We see that for every BER, using RD is more efficient than
not using RD. For BER$=10^{-5}$ and in several
cases when BER$=10^{-6}$ using $Rep$. even improves the Goodput
further. For example, in BER$=$$10^{-5}$ the Goodput
of $RD(25)$ is 780Mbps, compared to 600Mbps in the {\it No-RD} case.
With using $Rep$. the Goodput of $RD(25)$ is 860Mbps, over
$40\%$ improvement compared to {\it No-RD}. For BER$=$$10^{-7}$ and
BER$=0$ using $Rep$. decreases the performance for the reasons
mentioned previously.

\begin{figure}
\vskip 8cm
\includegraphics{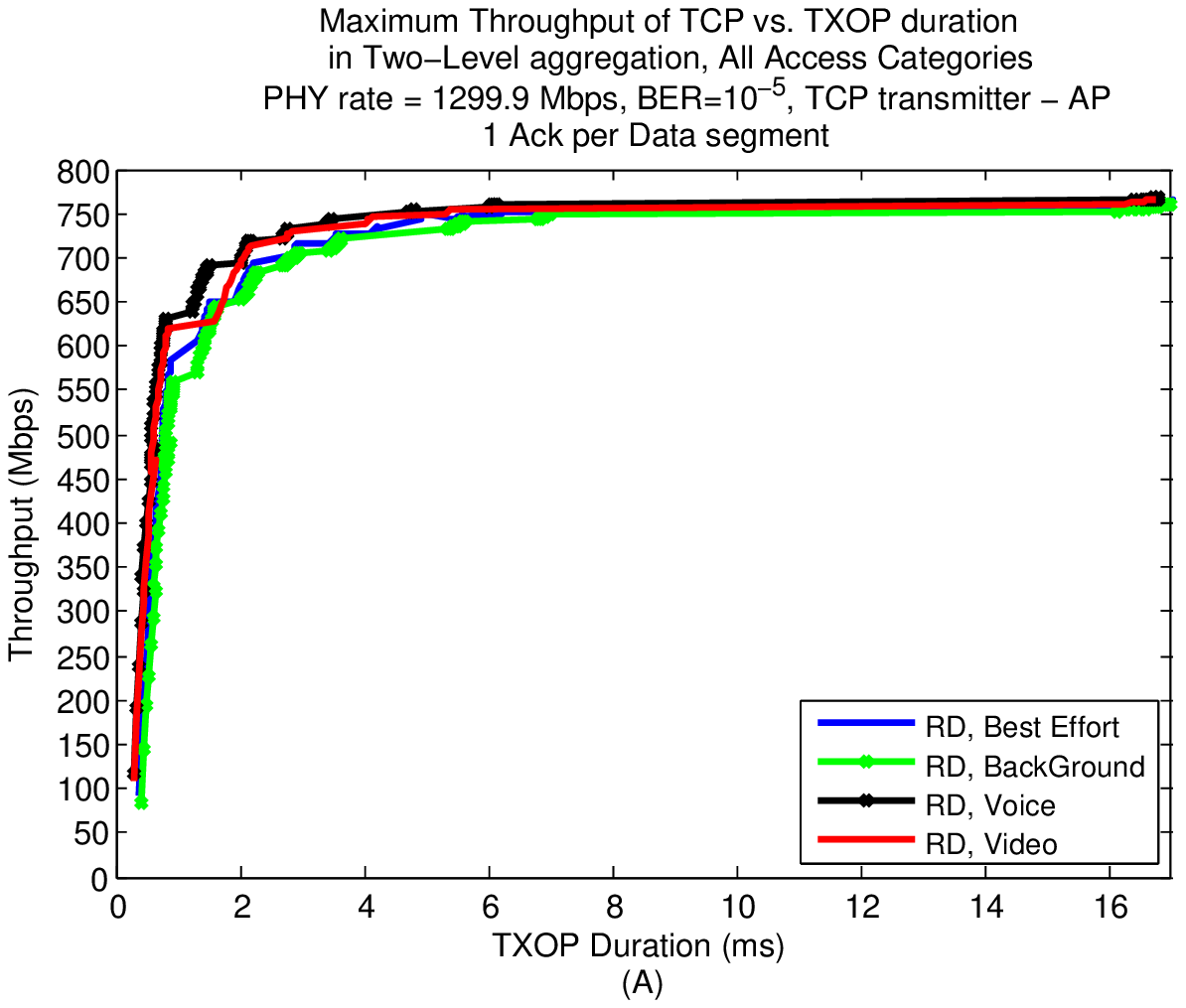}
\includegraphics{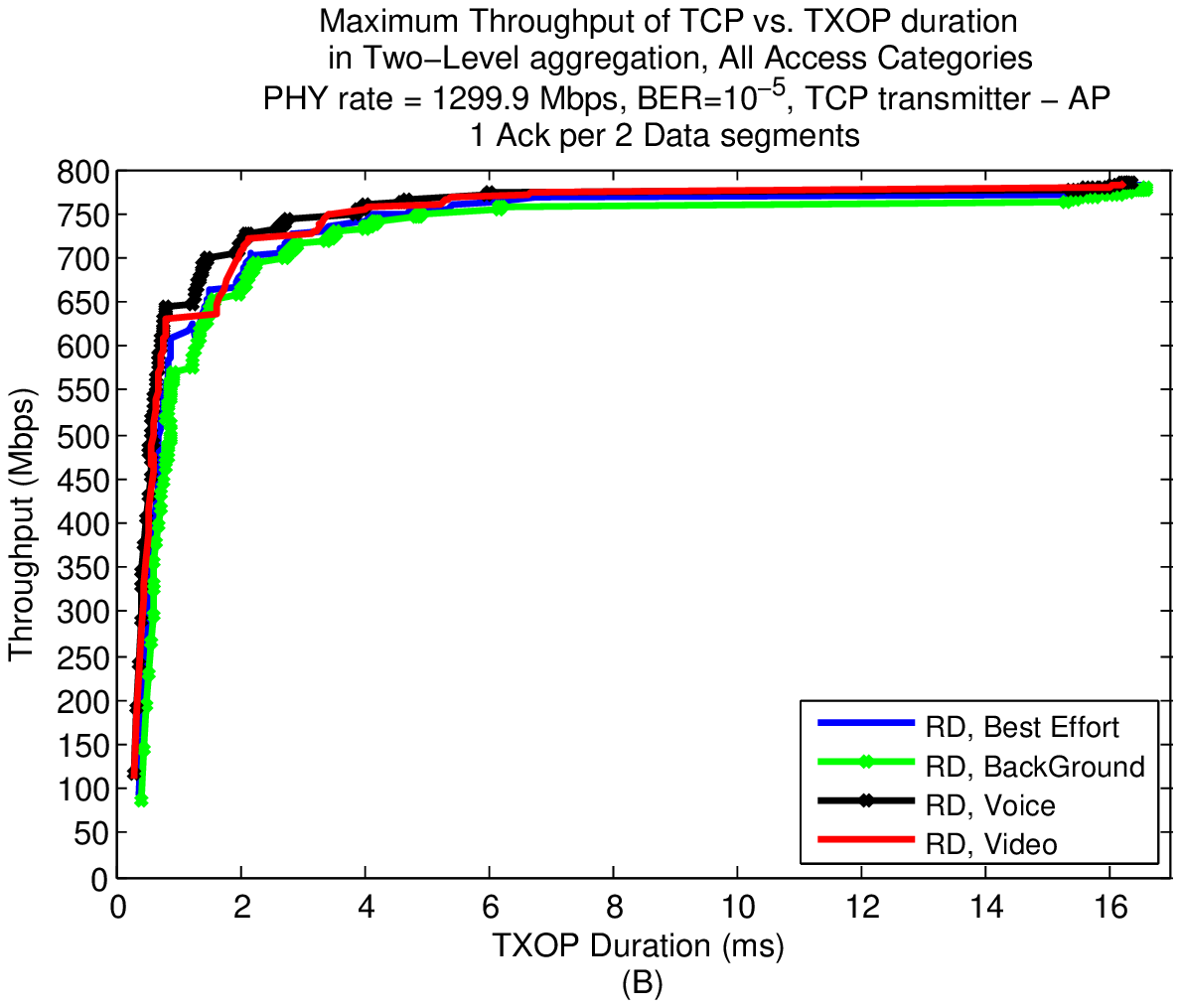}
\caption{The maximum Goodput vs. the duration of the TXOPs in the various ACs,
with and without TCP Delayed Acks, BER=0. }
\label{fig:txop1}
\end{figure}

Finally, in Figure~\ref{fig:txop1}  we show the maximum received
Goodput as a function of the TXOP for BER$=10^{-5}$. 
Recall that for BER$>$0 it 
is difficult to find the number of actually transmitted
MPDUs in every transmission of the TCP transmitter.
Therefore, we cannot use the same technique to compute
the maximum Goodput as in Section 4, BER$=$0.

Instead, we computed the average TXOP duration for $RD(1), RD(2),...,RD(25)$.
For every received Goodput $G$ we 
looked for the $RD(n), 1 \le n \le 25$ that
achieves $G$ with the shortest TXOP duration.

The outcomes and conclusions are similar in trend to those
in Figure~\ref{fig:txop} except that the achieved Goodputs are
much lower because of the positive BER. On the other hand the delays,
i.e. the length of the TXOPs, are shorter.  In BER$=$0 there
is no benefit to using TXOPs of more than $20 \mu s$
while for BER$=10^{-5}$ there is no benefit to using
TXOPs of more than $4 \mu s$.

\section{Summary}

This paper shows an example of the benefit achieved
when different layers in the protocol stack co-operate.
In particular, we show the improvement in the
TCP Goodput that is
achieved when the MAC layer of the IEEE 802.11ac standard
is aware of TCP traffic. Using Reverse Direction, the
contention between the TCP transmitter and receiver
is eliminated, and no time is wasted due to collisions.

Using also the Two-Level aggregation scheme, in an error-free channel
the TCP Goodput is improved by $20\%$ compared to contension 
based channel access. In an error-prone channel the TCP
Goodput is improved by $60\%$ also blindly using retransmission
of frames in A-MPDU frames.

This paper assumes only one TCP connection
in the system,
a scenario that is possible in small systems such as in the 
Home environment. A next research step is to investigate
the performance of Reverse Direction and aggregation
when the AP maintains several TCP connections at the
same time.

\section{Appendix}

In this Appendix we describe a Markov chain
model for the {\it No-RD} scheme and for
an error-free channel. The
Markov chain is based on  two assumptions : First, we assume that 
the case of 3 or more consecutive
collisions on the channel is very
rare. Notice that for the VO AC the probability
for 2 consecutive collisions is 
$(\frac{1}{4})^2(\frac{1}{8})^2 \sim 10^{-3}$. Therefore, we assume that
only two sizes of contention intervals
are used, $[0,...,CW_{min}-1]$ and $[0,...,2 \cdot CW_{min}-1]$.
Second, as already mentioned, we assume the saturated scenario where 
the TCP transmitter always has TCP Data segments to transmit
and that the TCP transmission window does not limit
the offered load.
In particular, we assume that for every $K_D, 1 \le K_D \le 64$,
the TCP transmitter can always transmit $K_D$ MPDUs
in a single transmission. 
Every MPDU contains 7 MSDUs of TCP Data segments.
The TCP receiver transmits all the TCP Acks it has
in one A-MPDU, up  to $178  \cdot 64$ in one transmission.

We also assume that every
TCP Ack acknowledges one TCP Ack.
The extension to the case of Delayed Acks is immediate.

We first present a Markov chain for the BE and BK ACs, which
are symmetrical in the sense that the AIFS of the AP and the
station are equal, Table~\ref{tab:model1}. We later show
what changes are needed for the VI and VO ACs that are
a-symmetrical.

\begin{figure}
\vskip 14cm
\includegraphics{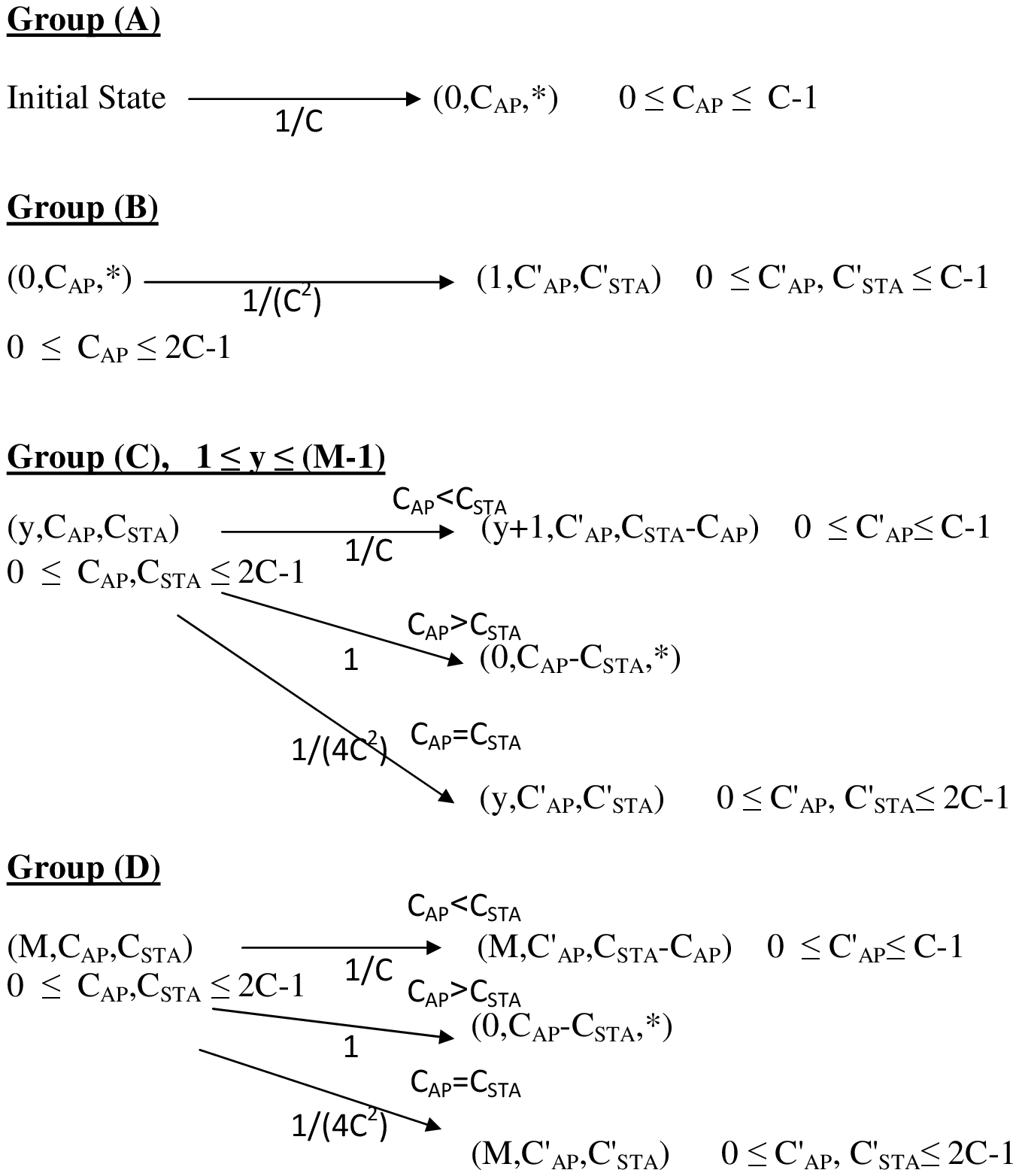}
\caption{Groups (A)-(D) of the Markov chain for the {\it No-RD} scheme and BE, BK ACs. $C$ stands for $CW_{min}$. '*' means there is no BackOff number.}
\label{fig:mrkv1}
\end{figure}

The Markov chain follows after the channel access state.
The set of its states, together with the transitions
among the states, is shown in Figure~\ref{fig:mrkv1}.
A state, except for the {\it Initial State}, represents 3 variables, and 
is denoted $(X,C_{AP},C_{STA})$.
$X$ denotes the number of $K_D \cdot 7$
TCP Acks that the TCP receiver accumulated
to transmit.
$C_{AP}$ and $C_{STA}$ 
are the values of the BackOff numbers of the AP and the station
respectively, to be used
in their next transmission attempt. These numbers
are multiplied by {\it SlotTime} to get the BackOff intervals.
When the station does not have TCP Acks to transmit,
it does not have a BackOff number, denoted by '*' in
Figure~\ref{fig:mrkv1}.

The Markov chain is based on 4 groups of states, denoted
Groups (A)-(D). In the first group, Group (A), there
is only one state, the {\it Initial State}. In this case the AP
randomly chooses its first BackOff number, and
move to the appropriate state in Group (B). Every
transition probability is $\frac{1}{C}$ where C stands
for $CW_{min}$ and notice that the BackOff number
is chosen at this stage from the interval $[0,...,C-1]$.

In Group (B) there are states of the form $(0,C_{AP},*), 
0 \le C_{AP} \le 2C-1$ . A state in this
group denotes that the station does not have
TCP Acks to transmit. 
The number of states in this group
is $2C$. The transitions
from every state in this group are according to the
randomly chosen $C_{AP}$ and $C_{STA}$, which
are chosen from the interval $[0,...,C-1]$.

In group (C) the station has $y \cdot K_D \cdot 7$
TCP Acks to transmit, $1 \le y \le M$. We explain
what is $M$ later.
Notice that there are three types
of transitions from a state in this group - a transition
when the AP transmits, when the station transmits and
when there is a collision. The transitions and their
corresponding probabilities are straight forward.
Notice that after the station transmits it is left
without TCP Acks, and the transition is to a state
in Group (B). We later explain why we assume that
the station is left without TCP Acks.

Notice that in principal the size of the Markov
chain is unlimited. We therefore look for a finite
size that will give analytical results within say 1\%
of those of the simulation. This seems to be a reasonable
error range. We therefore assume that the station cannot
accumulate more than $M \cdot K_D \cdot 7$ TCP Acks, and $M=20$
gives the desired error range.
Therefore, in every state in Group (C) the station
can transmit all the TCP Acks it has in one transmission.
(The station can accumulate up to $25 \cdot 64 \cdot 7$ TCP Acks
and transmit them in one A-MPDU ).

Group (D) of states is similar to Group (B),
except that the station already has
$M \cdot K_D \cdot 7$ TCP Acks to transmit,
and every another A-MPDU that the AP transmits is
dropped by the station.

We attach a $Time$ metric to every state. The $Time$ metric denotes
the time elapsed on the channel in this state.
The $Time$ metric of the {\it Initial State} is 0.
A state in which the AP transmits has a $Time$ metric
equals to $AIFS+BO+Preamble+T(DATA)+SIFS+BAck$.
{\it T(DATA)} is the transmission time
of an A-MPDU frame containing $K_D$ MPDUs of 7 TCP Data segments each.
A state in which the station transmits has a $Time$ metric
equals to $AIFS+BO+Preamble+T(ACK)+SIFS+BAck$.
{\it T(ACK)} is the transmission time
of an A-MPDU frame containing all the TCP Acks that the station has.
For a state in which there is a collision
the $Time$ metric equals to $AIFS+BO+Preamble+T(COL)+SIFS+Ack$.
$T(COL)$ is the maximum between $T(DATA)$ and $T(ACK)$
of the frames involved in the collision.
We denote by $Ts$ the Time
metric of state $S$.

We also attach a $Goodput$ metric to every state. Recall
that we consider a transmission of a TCP Data segment
to be a successful one only when a TCP Ack segment
is received for this segment.
Thus, the $Goodput$ metrics of the {\it Initial State}, every
state in which the AP transmits and every state that denotes a collision 
are all 0. For any other state the $Goodput$ metric
is the amount of bits of $X \cdot K_D \cdot 7$ TCP Data segments,
where $X$ is the number of $K_D \cdot 7$ TCP Acks
that are transmitted in the state, divided
by the $Time$ metric of the state. 
We denote by $Gs$ the Goodput 
metric of state $S$.

The Goodput $G$ of the system is
$G=
\frac{\sum_{s \in states} \pi s \cdot Ts \cdot Gs}
{\sum_{s \in states} \pi s \cdot Ts}$
where $\pi s, Ts$ and $Gs$ are the stationary probability, $Time$
metric and $Goodput$ metric respectively of every state $S$ in
the Markov chain.

\begin{figure}
\vskip 14cm
\includegraphics{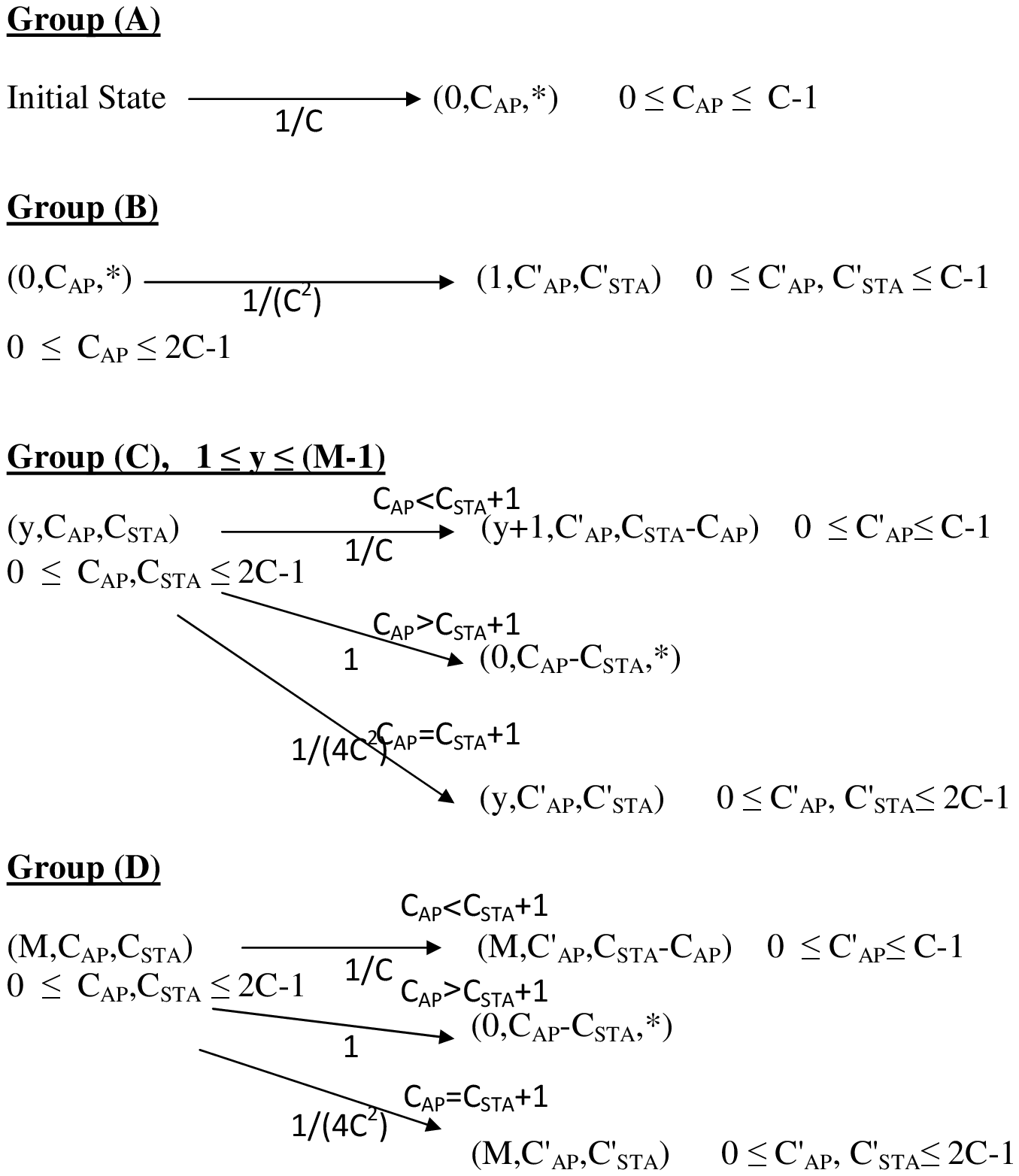}
\caption{Groups (A)-(D) of the Markov chain for the {\it No-RD} scheme and VI, VO ACs. $C$ stands for $CW_{min}$. '*' means there is no BackOff number.}
\label{fig:mrkv2}
\end{figure}

Concerning the VI and VO ACs, the AIFS of the
AP is shorter than that of the station by one SlotTime, Table~\ref{tab:model1}.
Therefore, a collision occurs when $C_{AP} = C_{STA}+1$,
the AP transmits when $C_{AP} < C_{STA}+1$ and the 
station transmits when $C_{AP} > C_{STA} +1$. The modified Markov
chain for these ACs is shown in Figure~\ref{fig:mrkv2}.

\clearpage


\bibliographystyle{abbrv}
\bibliography{main}


\end{document}